\documentclass[final]{siamltex}
\usepackage{amsmath,amsfonts,amssymb,epsfig} %

\def\define{\triangleq}

\def\Rplus{\mathbb{R}^{+}}
\def\R{\mathbb{R}}
\def\sD{{\cal D}}
\def\Utilde{\widetilde{U}}
\def\U{{U_p}}

\def\sf{{\cal F}}
\def\sS{{\cal S}}
\def\P{\mathbb{P}}
\def\E{\mathbb{E}}

\def\sD{{\cal D}}
\def\ind{\mathbb{I}}

\def\Sv{{\mathbb S_v}}
\def\Su{{\mathbb S_u}}
\def \NT {\mathbb{NT}}
\def \NTpm{\mathbb{NT^{\pm}}}
\def \NTp{\mathbb{NT^{+}}}
\def \NTm{\mathbb{NT^{-}}}

\def \Bpm{\mathbb{B^{\pm}}}
\def \Bp{\mathbb{B^{+}}}
\def \Bm{\mathbb{B^{-}}}

\def \Spm{\mathbb{S^{\pm}}}
\def \Sp{\mathbb{S^{+}}}
\def \Sm{\mathbb{S^{-}}}

\def \Kone{\mathbb{K}_1}
\def \etaT{\eta}

\def \NTu {\mathbb{\NTm}}
\def \Bu {\mathbb{\Bm}}
\def \Slu {\mathbb{\Sm}}

\def \NTv {\mathbb{NT}_v^{-}}
\def \Slv {\mathbb{S}_v^{-}}
\def \Bv {\mathbb{B}_v^{-}}

\def \X {{\tilde X}}
\def \Y {{\tilde Y}}

\def \zeroClosedTopen {{ [0,T]}}
\def \zeroClosedTclosed {{ [0,T]}}
\def \tClosedTclosed {{ [ t,T] }}
\def \tClosedTopen {{ [ t,T]}}
\def \zeroClosedTOpen {{ [ 0,T)}}

\newcommand{\cond}[1]{\big|\mathcal{F}_{#1}}

\def \EK {{\overline{ \mathcal{E}_K}}}
\def \GK {{\overline{ \mathcal{G}_K}}}

\newcommand{\e}[1]{\operatorname{e}^{#1}}

\def \abs#1{ \left \vert #1 \right \vert }

\newcommand{\dt}{\,\mathrm{d}t}
\newcommand{\du}{\mathrm{d}}

\newtheorem{Def}[theorem]{Definition}

\newtheorem{Cor}[theorem]{Corollary}

\newtheorem{Comparison Theorem}[theorem]{Comparison Theorem}
\newtheorem{remark}[theorem]{Remark}

\def\theequation{\arabic{section}.\arabic{equation}}

\title{Asymptotic Analysis for Optimal
Investment in Finite Time with
Transaction Costs}

\author{Maxim Bichuch%
\footnote{The author would like to thank Steven Shreve and Dmitry Kramkov  for useful comments and discussions.}\\ 
Department of Operations Research \& Financial Engineering\\
Princeton University\\
Princeton, NJ 08544\\
mbichuch@princeton.edu}

\begin{document}

\maketitle

\begin{abstract}
We consider an agent who invests in a stock and a money market account with the goal of maximizing the utility of his investment at the final time $T$ in the presence of a proportional transaction cost $\lambda>0$. The utility function is of the form $\U(c)=c^{p}/p$ for $p<1,~p\ne0$. We provide a heuristic and a rigorous derivation of the asymptotic expansion of the value function in powers of $\lambda^{1/3}.$ We also obtain a ``nearly optimal" strategy, whose utility asymptotically matches the leading terms in the value function.
\end{abstract}

\begin{keywords} 
Transaction costs, optimal control, asymptotic analysis, utility maximization
\end{keywords}
\begin{JEL} 
G13
\end{JEL} 
\begin{AMS}
90A09, 60H30, 60G44
\end{AMS}

\pagestyle{plain}

\pagestyle{myheadings}
\thispagestyle{plain}
\markboth{M. Bichuch}{Asymptotic Transaction Costs in Finite Time}

\setcounter{theorem}{0}
\setcounter{equation}{0}

\section{Introduction}
\setcounter{equation}{0} We consider the problem of an agent seeking to optimally invest in the presence of proportional transaction costs.  The agent can invest in a stock, modeled as a geometric Brownian motion with drift $\mu$ and volatility $\sigma$, and in a money market with constant interest rate $r$. The agent pays proportional transaction cost $\lambda>0$ for trading stocks, with the goal of optimizing the total utility of wealth at the final time $T$, when she would be required to close out her stock position and pay the resulting transaction costs. The utility function is given by $\U(c)\stackrel{\triangle}{=} \frac{c^p}{p}$, where $p<1, ~p\ne 0$. We refer to this optimized utility of wealth as the value function.
In this paper, we compute the asymptotic expansion of the value function up to and including the order $\lambda^\frac23.$ We also find a simple  {\em``nearly-optimal"} trading policy that, if followed, produces an expected utility of the final wealth that asymptotically matches the value function at the order of $\lambda^\frac23.$

In Section \ref{sec:setup} of this paper we define our model, state the HJB equation, and state Merton's result for the case of zero transaction costs. Under the smoothness assumption of the value function, in Section \ref{sec:heuristic} we provide a heuristic expansion of the value function in powers of $\lambda^{1/3}$. In the next section we use this heuristic expansion in order to build smooth functions $w^{\pm}$, which we later prove to be upper and lower bounds on the value function $u$. These functions also turn out to be sub- and supersolutions for the Hamilton-Jacobi-Bellman (HJB) equation. It is then possible to apply the Comparison Principle for viscosity solutions to conclude that the value function $u$, which is a viscosity solution for the HJB equation, has to be between the super- and subsolutions. However, this method is only applicable for $0<p<1$. Therefore we use a verification argument from stochastic calculus. In the final section we construct a simple policy and in Theorem \ref{thm:comp} prove that $w^{\pm}$ are indeed upper and lower bounds on the value function $u$. As a corollary we also get that the expected utility of the final wealth from the constructed policy is order $\lambda$ close to the value function, which makes this policy a {\em``nearly-optimal"} policy.

In the case of zero transaction cost, the agent's optimal policy is to keep a constant proportion of wealth $\theta$, which we call the {\em Merton proportion}, invested in stock. See Pham \cite{Pham}, or alternately the original paper of Merton \cite{Merton} where a solution to a similar investment and consumption problem with infinite time horizon appears.

When $\lambda>0$, the optimal policy is to trade as soon as the position is sufficiently far away from the {\em Merton proportion}. More specifically, the agent's optimal policy is to maintain her position inside a region that we refer to as the {\em``no-trade"} (NT) region. If the agent's position is initially outside the NT region, she should immediately sell or buy stock in order to move to its boundary. The agent then will trade only when her position is on the boundary of the NT region, and only as much as necessary to keep it from exiting the NT region, while no trading occurs in the interior of the region; see Davis, Panas, \& Zariphopoulou \cite{DavisPanasZariphopoulou}. Not surprisingly, the width of the NT region depends on time, which makes it difficult to pinpoint exactly the optimal policy. Moreover, the NT region degenerates when the {\em Merton proportion} $\theta=1$, i.e. it is optimally to be fully invested in stock, since in this case, the agent only needs to trade at the initial time to buy stock, and the final time to liquidate his position. We will not consider this case. The approach of this paper, is to expand the value function into a power series in powers of $\lambda^\frac13$. This approach, which leads to explicit results, was pioneered by Jane\v{c}ek \& Shreve \cite{JanecekShreve} in solving the infinite horizon investment and consumption problem. 
Many other papers have used asymptotical expansion including Goodman \& Ostrov \cite{GoodmanOstrov}, who showed how the first term in asymptotical expansion of the value function relates to a free boundary problem that minimizes a cost function. They also showed that the quasi-steady state density of the portfolio is constant in the NT region. 
Jane\v{c}ek \& Shreve \cite{JanecekShreve1}  used it to solve a problem of optimal investment and consumption with one futures contract, and Bichuch \cite{Bichuch} applied it to the case of two correlated futures contracts.
Dewynne, Howison, Law \& Lee \cite{DewynneHowisonLawLee} heuristsically found a time independent policy in a finite horizon problem with multiple correlated stocks. Under the assumption that {\em the principle of smooth fit} holds and that the boundaries are symmetrical around the Merton proportion, they heuristically  computed the asymptotic location of the boundaries of the NT region. We prove this result rigorously for a problem with one risky asset, and quantify the optimality of the proposed policy. Numerical results provided by Gennotte \& Jung \cite{GennotteJung} and Liu \& Loewenstein \cite{LiuLoewenstein} show that the optimal boundaries are not symmetrical around the Merton proportion and that they are complicated functions of time. For instance, Dai \& Yi \cite{DaiYi} find a time, of order $\lambda$ close to final time $T$, after which the agent would no longer buy stock. The intuitive explanation is that it is wasteful spending to buy extra stocks, standing very close to final time, only to sell them all a moment later, without realizing virtually any profit, since the agent held them for very little time. %
Our goal instead is to find a simple {\em``nearly-optimal"} policy. 
We rely on the results obtained by Dai \& Yi \cite{DaiYi}, who use a PDE approach to problem to show a connection between the optimal investment problem and a double obstacle problem. Using the theory of the obstacle problem, they show that the value function is smooth (see Theorem 5.1 for exact formulation). They also characterize the behaviors of the free boundaries.

Transaction costs were introduced into Merton's model by  Magill \& Constantinides \cite{MagillConstantinides}. Their analysis of the infinite time horizon investment and consumption problem, despite being heuristic, gives an insight into the optimal strategy and the existence of the NT region. A more rigorous analysis of the same infinite time horizon problem was given by Davis \& Norman \cite{DavisNorman}, who under certain assumptions showed that the value function is smooth. The viscosity solution approach to that infinite time horizon problem was pioneered by Shreve \& Soner \cite{ShreveSoner}, who significantly weakened the assumptions of Davis \& Norman \cite{DavisNorman}.

An alternative to the dynamic programming approach above is to use the martingale duality approach. Cvitani\v{c} \& Karatzas \cite{CvitanicKaratzas} in a finite time horizon investment problem using duality proved the existence of an optimal strategy, under the assumption that a dual minimization
problem admits a solution. Later Cvitani\v{c} \& Wang \cite{CvitanicWang} proved the existence of a solution to the dual problem.
In a more general framework with multiple assets Kabanov \cite{Kabanov99} proved the existence of an optimal strategy, also assuming the existence of a minimizer to the dual problem. Subsequent existence results under more relaxed assumptions were proved in Deelstra, Pham \& Touzi \cite{DeelstraPhamTouzi} and Campi \& Owen \cite{CampiOwen}.

While the problem of optimal investment in the presence of transaction costs is important in its own right, it has further value in the study of contingent claim pricing. Hodges and Neuberger \cite{HodgesNeuberger} proposed to price an option so that a utility maximizer is indifferent between either having a certain initial capital for investment or else holding the option but having initial capital reduced by the price of the option. This produces both a price and a hedge, the latter being the difference in the optimal trading strategies in the problem without the option and the problem with the option. This utility-based option pricing is examined in \cite{Bouchard}, \cite{ConstantinidesZariphopoulou99}, \cite{ConstantinidesZariphopoulou01}, \cite{DavisPanasZariphopoulou}. A formal asymptotic analysis of such an approach appears in Whalley \& Wilmott \cite{WhalleyWilmott}. They assume a power expansion for the value function and compute the leading terms of it for both the case of holding the option liability and the case without it. Their proof corresponds to the heuristic derivation section in this paper. We believe this paper is a step in the direction of providing a rigorous proof to a corresponding result with power utility.  %

\section{Model set-up and known results} \label{sec:setup}
\setcounter{equation}{0}

The set-up of the model is similar to Shreve \& Soner \cite{ShreveSoner}, only with finite time horizon $T>0$. An agent is given an initial position of $x$ dollars in the money market and $y$ dollars in stock. The stock price is given by 
\begin{equation}
dS_t= \mu S_t dt+ \sigma S_t dW_t,~S_0=1,
\label{eq:S}
\end{equation}
where $\mu$ and $\sigma$ are positive constants and $\{W_t,t
\geq 0\}$ is a standard Brownian motion on a filtered probability
space $\bigl(\Omega,\mathcal F,\{\mathcal F_t\}_{0 \le t \le T},\P\bigr)$. We assume a constant positive interest rate $0<r<\mu$.  The agent must choose a policy consisting of two adapted processes $L$ and $M$ that are nondecreasing and right-continuous with left limits, and $L_{0-} =M_{0-} = 0$. $L_t$ represents the cumulative dollar value of stock purchased up to time $t$, while $M_t$ is the cumulative dollar value of stock sold.

Let $X_t$ denote the wealth invested in the money market and $Y_t$
the wealth invested in stock, with $X_{0-} = x$, $Y_{0-} = y$.
The agent's position evolves as
\begin{eqnarray}
 \du X_t &=& r X_t\dt - (1+\lambda)\,\du L_t + (1-\lambda)\,\du
 M_t, \label{eq:position1}\\
 \du Y_t &=& \mu Y_t \dt + \sigma Y_t \,\du W_t + \du L_t - \du
 M_t.\label{eq:position2}
\end{eqnarray}

The constant $\lambda \in (0,1)$ appearing in these
equations accounts for proportional
transaction costs, which are
paid from the money market account.

\begin{remark}
\label{remark:l.s.c}
From \eqref{eq:position1} and \eqref{eq:position2} it follows that $X_{\cdot}(\omega)+Y_{\cdot}(\omega)-\lambda\abs{Y_{\cdot}(\omega)}$ is a lower semi-continuous function.
\end{remark}

Define the \emph{solvency region}
$$
\Sv \triangleq
\left\{(x,y);\:x+(1+\lambda)
\,y > 0,\: x+(1-\lambda) \,y >
0\right\}.
$$
The policy $(L_s,M_s)\big|_{s\in\tClosedTclosed}$ is \emph{admissible} for the initial position $(t,x,y)$, if $(X_s,Y_s)$ starting from $(X_{t^-}, Y_{t^-})=(x,y)$ and given by (\ref{eq:position1}), (\ref{eq:position2}) is in $\overline{\Sv} $ for all $s\in\tClosedTclosed$. Since the agent may choose to immediately rebalance his position, we agree the initial time to be $t-$. We denote by
$\mathcal A(t,x,y)$ the set of all such policies. We note that
$\mathcal A(t,x,y) \neq \emptyset$ if and only if $(x,y) \in
\overline{\Sv}$.

We introduce the agent's \emph{utility function} $\U$ defined for
all $c \geq 0$ by $\U(c) \triangleq c^{p}/p$ for $p<1,~p\ne0$. (An analysis
along the lines of this paper is also possible for $U_0(c)=\log
c$, but we omit that in the interest of brevity.) For convenience we agree to treat $\U(0)=\frac{0^p}{p}\triangleq-\infty$ when $p<0$ here and in the rest of this paper. Define the value function as the supremum of the utility of the final cash position, after the agent liquidates her stock holdings
\begin{equation}
v_0(t,x,y)\define\sup_{(L,M)\in\mathcal A(t,x,y)} \E \left[ \U(X_T+Y_T-\lambda|Y_T|)\cond{t}\right],\quad (t,x,y)\in \zeroClosedTclosed\times\overline{\Sv}.
\label{eq:v}
\end{equation}
For $(t,x,y)\in\zeroClosedTclosed\times\overline{\Sv}$ and $\beta\ge0$ we also define an auxiliary value function
\begin{equation}
v_{\beta}(t,x,y)\define\sup_{(L,M)\in\mathcal A(t,x,y)} \E \left[e^{-\beta(T-t)} \U(X_T+Y_T-\lambda|Y_T|)\cond{t}\right].
\label{eq:v-beta}
\end{equation}
Clearly 
\begin{equation}
v_0(t,x,y)=e^{\beta(T-t)}v_{\beta}(t,x,y),~\beta\ge0,(t,x,y)\in\zeroClosedTclosed\times\overline{\Sv}.
\label{eq:v-v-beta}
\end{equation}
For the rest of this paper we will concentrate on finding $v_{\beta}.$ 

\begin{lemma} \label{lemma:zero_boundary}
{\rm
For $\beta\ge0$, $t\in\zeroClosedTclosed$ and $(x, y) \in \partial \Sv$, the only admissible policy is to jump
immediately to the origin and remain there.  In particular, $X_T=Y_T=0$, and $v_{\beta}(t,x,y)=0$ when $0<p<1$ and $v_{\beta}(t,x,y)=-\infty$ when $p<0.$
}
\end{lemma}

{\sc Proof:}
The proof of this Lemma is a modification of Remark 2.1 in Shreve \& Soner \cite{ShreveSoner}.

$\hfill\Box$

The problem with $\lambda=0$ is similar to the problem solved by Merton \cite{Merton}.  It can be easily seen that the optimal policy always keeps a
wealth proportion
\begin{equation}\label{2.a}
\theta = \frac{\mu -r }{(1-p)\sigma^2},
\end{equation}
in the stock, see Pham \cite{Pham}. We call $\theta$
the {\em Merton proportion}.  For $\lambda=0$,
\begin{equation}\label{2.b}
v_{\beta}(t,x,y)=\frac{1}{p}e^{pA(T-t)}(x+y)^{p},~\beta\ge0,(t,x,y)\in\zeroClosedTclosed\times\overline{\Sv},
\end{equation}
where
$
A \triangleq r-\frac{\beta}{p}+(\mu-r)\theta-\frac12(1-p)\sigma^2\theta^2=r-\frac{\beta}{p}+\frac12\frac{(\mu-r)^2}{(1-p)\sigma^2}.
$

Note, that $v_\beta|_{\zeroClosedTclosed\times\overline{\Sv}}<\infty, \beta\ge0$. This is clear in case of $\lambda=0$, and $v_\beta$ cannot increase as $\lambda$ increases. This is not the case in the infinite time horizon case, when a condition on the parameters is required to assure the finiteness of the value function. 
\begin{remark}\label{remark:beta}
Fix any $\beta \ge 0$ such that $pA <0$. For the rest of this paper we will deal with that fixed $\beta$ and for convenience we will drop the subscript and refer to the value function $v_\beta$ simply as $v$. It turns out that this $v=v_\beta$ is easier to find than $v_0$, but because of \eqref{eq:v-v-beta} there is no loss of generality in doing so. When $p<0$ the choice $\beta=0$ is suitable, however, the case when $0<p<1$ requires a strictly positive $\beta$.
The term $pA$ can be understood as the optimal growth rate in the sense of Akian, Menaldi \& Sulem \cite{AkianMenaldiSulem1}, and $\beta$ as the investor impatience.

\end{remark}
The following theorem is parallel to the one proved by Davis, Panas, \& Zariphopoulou \cite{DavisPanasZariphopoulou} and Shreve \& Soner \cite{ShreveSoner}.

\begin{theorem} \label{thm:HJB}
The value function $v(t,x,y)$ defined by \eqref{eq:v-beta} is a viscosity solution of the following HJB equation (\ref{eq:HJB}) on $\zeroClosedTopen\times\Sv$:
\begin{equation} \label{eq:HJB}
\min\left\{-v_t+\mathcal L v,\, -(1-\lambda)v_x +
v_y,\,(1+\lambda)v_x - v_y\right\}=0,
\end{equation}
where the second-order differential
operator $\mathcal L$ is given
by
\begin{equation}
(\mathcal L v ) (t,x,y)
\define - \frac{1}2 \sigma^2
y^2 \,v_{yy}(t,x,y)- \mu y \,v_y(t,x,y)
- rx \,v_x(t,x,y)+\beta v(t,x,y),
\label{eq:L}
\end{equation}
together with the terminal condition 
\begin{equation}
v(T,x,y)=\U(x+y-\lambda|y|),~(x,y)\in\overline\Sv.
\label{eq:termnina_v}
\end{equation}
\end{theorem}

Power utility functions lead to \emph{homotheticity} of the value function: for $\gamma>0$,
\begin{equation}\label{2.6}
 v(t, \gamma x, \gamma y)
=\gamma^{p} \,v(t, x,y), \quad (t,x,y)\in\zeroClosedTclosed\times\overline \Sv.
\end{equation}
This is because $(L_s,M_s)\big|_{s\in\tClosedTclosed} \in \mathcal A(t, x,y)$ $\Leftrightarrow$
$(\gamma L_s,\gamma M_s)\big|_{s\in\tClosedTclosed} \in \mathcal A(t, \gamma x,\gamma y)$.
Consequently, the problem reduces to that of two variables. With $\Su \define
\left(-1/\lambda,1/\lambda\right)$, we define
\begin{equation} \label{eq:u}
u(t, z) \define v(t,1-z,z),\qquad (t,z) \in \zeroClosedTclosed\times\overline\Su.
\end{equation}
In other words, we make the change of variables $z = y/(x+y)$, $1-z = x/(x+y)$, which maps the solvency region $\Sv$ onto the interval $\Su$. Then
\begin{equation} \label{eq:valueFct}
v(t,x,y) = (x+y)^{p}\, u\biggl(t, \frac{y}{x+y}\biggr),\quad
(t,x,y)\in\zeroClosedTclosed\times\Sv.
\end{equation}

The counterpart to Theorem \ref{thm:HJB} for the reduced-variable function $u$ is the following lemma. It is parallel to Proposition 8.1 from Shreve \& Soner \cite{ShreveSoner}.

\begin{lemma}\label{lemma:reduction}
On $\zeroClosedTopen \times\Su$, $u$ is a viscosity solution of the HJB equation $\mathcal H(u)=0,$ where
\begin{eqnarray}  
&&\!\!\!\!\!\!\!\!\!\!\!\!\!\!\mathcal H(u)\define\min\Bigl\{ -u_t(t,z)+\mathcal D u(t,z),
\:\lambda pu(t,z)+(1-\lambda z)\,u_z(t,z)\Bigr.\label{eq:HJB2}\\
&&\quad\Bigl.\lambda pu(t,z)-(1+\lambda z)
\,u_z(t,z)\Bigr\} ,\nonumber
\end{eqnarray}
\begin{equation} \begin{split} 
(\mathcal D u)(t,z) \define \:&
-p\Bigl(A -\frac{1}2 \sigma^2(1-p)(z-\theta)^2\Bigr)\,u(t,z)
\\ & +
(1-p)\sigma^2z(1-z)(z-\theta)\,u_z(t,z)-\frac{1}2 \sigma^2
z^2(1-z)^2 u_{zz}(t,z),
\end{split} 
\label{eq:Du}
\end{equation}
with the terminal condition $u(T,z)=\U(1-\lambda|z|),~z\in\overline\Su$.
\end{lemma}

For future convenience for $(t,z)\in\zeroClosedTclosed\times\Su$ we also define two first-order differential operators
\begin{eqnarray}
(\mathcal B u)(t,z)&\define&\lambda pu(t,z)-(1+\lambda z)u_z(t,z),\label{eq:Bu}\\
(\mathcal S u)(t,z)&\define&\lambda pu(t,z)+(1-\lambda z)u_z(t,z).\label{eq:Su}
\end{eqnarray}

Dai \& Yi \cite{DaiYi} show that the optimal policy can be described in
terms of two functions $0 \le z_1(t)< z_2(t)\le \infty$
which define the {\em``no-trade"} region as a function of $t$ 
$$\left\{(t,x,y)\Big| t\in\zeroClosedTopen, (x,y)\in\Sv,z_1(t)<\frac{y}{x+y}<z_2(t)\right\}.$$
In this region $-v_t+\mathcal L v$ is zero. Here and in the rest of this paper, the derivative with respect to $t$ at $t=0$ or $t=T$ should be understood as the right-sided or left-sided derivative respectively. Moreover, if the second derivative with respect to $z$ does not exist, then the desired property should be satisfied with both one-sided second derivatives. 
If $Y_t/(X_t+Y_t)<z_1(t)$ one should buy stock in order to bring this
ratio to the boundary $y/(x+y)=z_1(t)$ of the {\em ``no-trade"} region. In this region $(1+\lambda) v_x-v_y$ is zero. If $Y_t/(X_t+Y_t)>z_2(t)$ one should sell stock in order to bring this ratio to the other boundary $y/(x+y)=z_2(t)$ of the {\em ``no-trade"} region. In this region $v_y-(1-\lambda) v_x$ is zero; see Davis, Panas, \& Zariphopoulou \cite{DavisPanasZariphopoulou} and Shreve \& Soner \cite{ShreveSoner}. 

\section{Heuristic derivation by Taylor series}
\setcounter{equation}{0} \label{sec:heuristic}

In this section we derive several terms of a power series expansion of the value function by a heuristic method. Similar to Shreve \& Soner \cite{ShreveSoner} and Jane\v{c}ek \& Shreve \cite{JanecekShreve}, we will assume that the {\em``no-trade"} region in the reduced variable form is
\begin{equation}
\NT=\left\{ (t,z) \Big| t\in\zeroClosedTopen,z \in (z_1(t), z_2(t))\right\},
\label{eq:NT}
\end{equation} 
and that $|z_i(t)-\theta|=O(\lambda^{\frac13}),~i=1,2$.

\begin{remark}\label{remark:order}
In the line above and for the rest of this paper, we have used the following standard notation: 

For a function $f$ defined on $\mathbb D\times\mathbb D_1\subset\Rplus\times\R^2$ we say that $f(\lambda,t,z)=O(\lambda^q),~q>0$ if there exist a constant $C$ independent of $(t,z)$ such that
\begin{equation}
\abs{ f(\lambda,t,z)} \le C \lambda^q,~\forall (t,z)\in\mathbb D_1
\label{eq:O}
\end{equation}
for all $\lambda >0$ small enough.

We say that $f(\lambda,t,z)=o(\lambda^q),~q>0$ if \eqref{eq:O} is true for any $C>0$.
Similar definition can be made if $f$ is just a function of $\lambda $ and one additional variable.

To be even more precise, in either case, we will allow $C$ to depend only on the constants $\mu, r, \sigma, T, p, \beta$, unless noted otherwise.

\end{remark}

We believe that $\NT$ has the form \eqref{eq:NT} for all times $t$ except those ``very close" to $T$. Intuitively
a change in strategy for times $O(\lambda)$ close to $T$ will affect the expected utility of the final wealth only at order $O(\lambda)$, since buying an extra stock and holding it $O(\lambda)$ time only affect wealth at $O(\lambda)$. However we can neglect this effect, since we are only looking to find the value function up to the order of $O(\lambda^{\frac23})$. It is not hard to see that $v$ is continuous on $\zeroClosedTclosed\times\overline{\Sv}$ and in this paragraph we will also assume that $v \in C^{1,2,2}(\zeroClosedTopen\times\Sv)$. It follows that $u \in C^{1,2}(\zeroClosedTopen\times\Su)$. Moreover, for $t\in\zeroClosedTopen$ we will assume that
\begin{eqnarray}
\lambda pu(t,z)-(1+\lambda z)u_z(t,z) &=&
0,\quad -\frac{1}{\lambda}<z\leq z_1(t),\label{eq:HJBp1}\\
-u_t(t,z)+\mathcal D u(t,z)&=&
0,\quad z_1(t)\leq z\leq z_2(t),\label{eq:HJBp2}\\
\lambda pu(t,z)+(1-\lambda z)u_z(t,z) &=&0,\quad
z_2(t)\leq z<\frac{1}{\lambda}.\label{eq:HJBp3}
\end{eqnarray}

Equations (\ref{eq:HJBp1}) and (\ref{eq:HJBp3}) are consequences of the directional derivative of $v(t,x,y)$ being zero in the directions of transaction in the
regions in which it is optimal to buy stock and to sell stock, respectively. These equations imply for $t \in \zeroClosedTopen$ that
\begin{eqnarray}
u(t,z) &=& u(t,z_1(t))\left(\frac{1+\lambda z}{1+\lambda z_1(t)}
\right)^{p},\quad -\frac{1}{\lambda}<z\leq z_1(t),
\label{eq:SolutionBS}\\
u(t,z) &=& u(t,z_2(t))\left(\frac{1-\lambda z}{1-\lambda z_2(t)}
\right)^{p},\quad z_2(t)\leq z<\frac{1}{\lambda}.
\label{eq:SolutionSS}
\end{eqnarray}

There is no explicit solution to the free boundary problem (\ref{eq:HJBp1}) - (\ref{eq:HJBp3}). We thus assume that in the $\NT$ region $u(t,z)$ has an expansion around the value function with zero transaction costs in powers of $\lambda^{1/3}$, and we expect the coefficient of $\lambda^{1/3}$ to be zero.  In order to work with this expansion, we need to also include the variable $z$, and we do that using powers of $z-\theta$. For $(t,z)\in \overline\NT$ we assume
\begin{equation} \label{eq:uTaylor}
\begin{split}
  u(t,z) & = \gamma_0(t) - \gamma_1(t) \,\lambda^\frac{1}3 - \gamma_2(t)\, \lambda^\frac{2}3
  - \gamma_3(t) \,\lambda - \gamma_{40}(t)\,\lambda^\frac{4}3 - \gamma_{41}(t)\,(z-\theta)
  \lambda \\
  &\quad - \gamma_{42}(t)\,(z-\theta)^2 \lambda^\frac{2}3 -
  \gamma_{43}(t)\,(z-\theta)^3 \lambda^\frac{1}3 - \gamma_{44}(t)\,(z-\theta)^4 +
  O\bigl(\lambda^\frac{5}3\bigr).
\end{split}
\end{equation}

We can now compute and equate the derivatives of $u$ with respect to $z$ across the boundaries of the $\NT$ region, similar to what is done in Jane\v{c}ek \& Shreve \cite{JanecekShreve}, Section 3, {\emph {Heuristic derivation by Taylor series}}. For sake of brevity this computation is omitted. The result is that for $(t,z)\in \overline \NT$

\begin{eqnarray}
u(t,z) &=& \frac{1}{p}e^{pA(T-t)} -\gamma_2(t) \,\lambda^\frac{2}3 -\gamma_3(t)\lambda \label{eq:heuristic-u}\\
&&-\frac{e^{pA(T-t)}}{\nu}\left[ \frac32(z-\theta)^2\lambda^{\frac23} -\frac{(z-\theta)^4}{\nu^2}\right]+O(\lambda^{\frac53}),\nonumber
\end{eqnarray}
where the coefficient $\gamma_3(t)$ is irrelevant for the rest of this paper and 
\begin{eqnarray}
\gamma_2(t) &\define& \left(\frac{9}{32} (1-p)
\theta^4(1-\theta)^4\right)^\frac{1}3 (T-t)e^{pA(T-t)}
\sigma^2,~t\in\zeroClosedTclosed,\label{eq:gamma-zeta}\\
\nu&\define& \left(
\frac{12}{1-p}\theta^2(1-\theta)^2\right)^\frac{1}3.\label{eq:nu}
\end{eqnarray}
For convenience we also define the constant 
\begin{equation}
\gamma_2\define\left(\frac{9}{32} (1-p)
\theta^4(1-\theta)^4\right)^\frac{1}3\sigma^2 
\label{eq:gamma2}
\end{equation}
so that we can write $\gamma_2(t)=\gamma_2e^{pA(T-t)}(T-t).$ 

\begin{remark}\label{remark:heuristic_res}
The heuristic method and the results above are very similar to the ones in Jane\v{c}ek \& Shreve \cite{JanecekShreve}, and the method is essentially similar to the one in Whalley \& Wilmott \cite{WhalleyWilmott}. It should not come as a surprise that even though $\gamma_3(t)$ is not important, but, for example, $\gamma_{42}(t)$ is, since the later term would add a contribution of order $\lambda^{\frac23}$ in $u_{zz}$. Also notice that in case $\theta=0$ ($\mu=r$) or $\theta=1$, i.e. the agent is not invested in stock at all, or is fully invested, there is no loss of the value function at the order of $\lambda^\frac23$, because these positions do not require trading except possibly at the initial and final times, so the loss will only be at the order of $\lambda$ and $\gamma_2=0.$  As previously stated, we exclude these two cases.

\end{remark}

\section{Rigorous asymptotic expansion}
\setcounter{equation}{0} \label{sec:expansion}

In this section we build the functions $w^{\pm}$ and prove in Theorem \ref{thm:comp} that they are tight lower and upper bounds on the value function $u$. They also turn out to be  sub- and supersolutions of the HJB equation; see \cite{CL}, \cite{CEL}, \cite{CrandallIshiiLions}. We have already stated the first classical Theorem \ref{thm:HJB} and its corollary Lemma \ref{lemma:reduction}, asserting that the value function is a viscosity solution of the HJB equation. One way to proceed to establish that supersolutions and subsolutions are indeed upper and lower bounds on the value function is to use a comparison theorem. Theorem 8.2 from Crandall, Ishii \& Lions \cite{CrandallIshiiLions} asserts that any supersolution dominates any subsolution. Since the value function is both a viscosity sub- and super solution, the desired result would follow. However, a standard comparison theorem requires finite boundary values. In our case, that means that it can be applied only when $0<p<1$ and the value function is zero on $\zeroClosedTclosed\times\partial \Su$. In the case $p<0$ it cannot be applied since the value function is $-\infty$ on the boundary of the solvency region. Therefore, similar to  Jane\v{c}ek \& Shreve \cite{JanecekShreve}, we instead choose to use a version of the verification lemma from stochastic calculus that can be applied to both cases; see Theorem \ref{thm:comp}.

The main theorem of this paper is:

\begin{theorem}\label{thm:main}
Assume $p<1,~p\ne0,~pA<0~,\theta>0,~\theta\ne1$ and $\lambda >0.$ Fix $\Kone \subset \R$ a compact. For $\lambda>0$ small enough such that $\Kone\subset\Su$, and for $(t,z) \in \zeroClosedTclosed\times\Kone$, the value function satisfies
\begin{equation}
u(t,z) = \frac{1}{p}e^{pA(T-t)}- \Bigl(\frac{9}{32}(1-p)\,\theta^4(1-\theta)^4\Bigr) ^{\frac{1}3} (T-t)e^{pA(T-t)}\,\sigma^2\,\lambda^{\frac23} + O(\lambda),
\end{equation}
where the remainder $O(\lambda)$ holds independently of $(t,z),$ but depends on the compact $\Kone$. 
Moreover, there exist a simple strategy $(\tilde L, \tilde M)$, constructed in Lemma \ref{lemma:existence}, which is ``nearly optimal". That is, for $(t,z) \in \zeroClosedTclosed\times\Kone$, the expectation of the discounted utility of the final wealth for this strategy satisfies
\begin{eqnarray*}
&&\E \left[e^{-\beta(T-t)} \U\left(\X_T+\Y_T-\lambda\abs{\Y_T}\right)\cond{t}\right] \\
&&=\frac{1}{p}e^{pA(T-t)}-\Bigl(\frac{9}{32}(1-p)\,\theta^4(1-\theta)^4\Bigr) ^{\frac{1}3} (T-t)e^{pA(T-t)}\,\sigma^2\,\lambda^{\frac23} + O(\lambda),
\end{eqnarray*}
where $(\X_s, \Y_s)\big|_{s\in\tClosedTclosed}$, is the diffusion associated with this trading strategy. In other words, it matches the value function $u$ at the order of $\lambda^{\frac23}$. Here again the term $O(\lambda)$ holds independently of $(t,z),$ but depends on the compact $\Kone$.
\end{theorem}

However, first we need to prove an auxiliary theorem:

\begin{theorem}\label{Thm1}
Assume $p<1,~p\ne0,~pA<0~,\theta>0,~\theta\ne1$ and $\lambda >0.$ Then there exist four smooth functions $\delta_i^{\pm} \in C^2(\zeroClosedTclosed),~i=1,2$, defined in Lemma \ref{lemma:roots}, additionally, there exist two continuous functions $w^{\pm}(t,z) \in %
C^{1,1}(\zeroClosedTopen\times \Su)$ with the following properties. The functions $w^{\pm}$ are twice continuously differentiable with respect to $z$ in $\zeroClosedTopen\times \Su$ except on the curves $(t,\theta+\delta_i^{\pm}(t)),~t\in\zeroClosedTclosed,~i=1,2.$ On these curves $w^{\pm}$ have one-sided limits of their second derivatives. Moreover, they satisfy $\pm (\mathcal H)(w^{\pm}) \ge 0$ on $\zeroClosedTopen\times \Su,$ where on curves $(t,\theta+\delta_i^{\pm}(t)),~t\in\zeroClosedTclosed, ~i=1,2,$ the second derivative with respect to $z$ can be either one of the one-sided derivatives. In addition, $w^{\pm}$ satisfy the boundary condition $w^{\pm}(t,z)=0$ if $0<p<1$ and $w^{\pm}(t,z)=-\infty$ if $p<0$ for $(t,z)\in \zeroClosedTclosed\times \partial \Su$, and the final time condition inequality $\pm w^{\pm}(T,z) \ge \U(1-\lambda \abs{z}),~z\in\overline\Su$. In addition, we have $w^{\pm}(t,\theta) = \frac{1}{p} e^{pA(T-t)} -\gamma_2(t)\lambda^\frac{2}3+O(\lambda),~t\in\zeroClosedTclosed$.
\end{theorem}

The plan is then to rigorously argue that for $(t,x,y)\in\zeroClosedTclosed\times\Sv$ and for any admissible trading strategy $(L_s,M_s)\big|_{s\in\tClosedTclosed}\in\mathcal A(t,x,y)$ with the corresponding diffusion $(X_s, Y_s)\big|_{s\in\tClosedTclosed}$ starting from $(X_{t-}, Y_{t-})=(x,y)$ and given by \eqref{eq:position1} and \eqref{eq:position2} that $(X_s+Y_s)^pw^{+}(s,Y_s/(X_s+Y_s))$ is a supermartingale. Using the fact that $w^{+}(T,z) \ge \U(1-\lambda \abs{z}),~z\in\overline\Su$, it follows that
\begin{eqnarray}
(x+y)^pw^{+}(t,y/(x+y))&\ge& \E \left[(X_T+Y_T)^pw^{+}(T,Y_T/(X_T+Y_T))\cond{t}\right] \label{eq:supermart}\\
&\ge& \E \left[\U(X_T+Y_T-\lambda \abs{Y_T})\cond{t}\right].\nonumber
\end{eqnarray}
Taking supremum over all admissible strategies and dividing by $(x+y)^p$, it follows that $w^{+}(t,y/(x+y)) \ge u(t,y/(x+y)).$

For the other direction, we would need to find a ``nearly-optimal" policy \\
$(\tilde L_s,\tilde M_s)\big|_{s\in\tClosedTclosed}\in\mathcal A(t,x,y)$ with the corresponding diffusion $(\X_s, \Y_s)$ for $s\in\tClosedTclosed$ starting from $(\X_{t-}, \Y_{t-})=(x,y)$, such that $(\X_s+ \Y_s)^{p}w^{-}(s,\Y_s/(\X_s+\Y_s))$ is a submartingale. Using the fact that $w^{-}(T,z) \le \U(1-\lambda \abs{z}),~z\in\overline\Su$, it follows that
\begin{eqnarray}
&&(x+y)^{p}w^{-}(t,y/(x+y))\le \E \left[(\X_s+ \Y_s)^{p}w^{-}(T,\Y_T/(\X_T+\Y_T))\cond{t}\right] \label{eq:submart}\\
&&\le \E \left[\U\left(\X_T+\Y_T-\lambda \abs{\Y_T}\right)\cond{t}\right] \le v(t,x,y).\nonumber
\end{eqnarray}
Dividing by $(x+y)^{p}$ we conclude that $w^{-}(t,y/(x+y)) \le u(t,y/(x+y)).$ Hence $w^{-}\le u\le w^{+}$ on $\zeroClosedTclosed\times\Kone$.

Finally, because $w^\pm(t,z) = w^\pm(t,\theta) + O(\lambda)$  and because $w^{+}(t,z) = w^{-}(t,z) + O(\lambda)$ for $(t,z)\in\zeroClosedTclosed\times\Kone$, we will conclude that $u(t,z) = w^\pm(t,\theta) + O(\lambda) = u(t,\theta)+O(\lambda)$. 
In Theorem \ref{thm:comp} it will also be shown that the expected utility of the ``nearly-optimal" policy, which is defined in Section \ref{sec:nearly-optimal_strat} is bounded below by $w^{-}$. We make the above heuristic arguments precise in Theorem \ref{thm:comp}.

\vspace{\baselineskip}
{\sc Proof:}
The proof of Theorem \ref{Thm1} is divided into five steps:  
\subsection{ \emph{{Step 1: The $\NT$ region and other sub-regions of $\Su$}}} 

We recall $\gamma_2(t), \gamma_2$ and $\nu$ of (\ref{eq:gamma-zeta}),  \eqref{eq:nu} and \eqref{eq:gamma2} respectively. Set $\xi(t) \define  \sqrt{\frac{2}3 p(T-t)\gamma_2+B},~t\in\zeroClosedTclosed$, where we set $B\define\frac{2}3 \abs{p}T\gamma_2+1$, chosen to make $\xi$ well defined. 
We next define
\begin{equation}
h(\delta)\define\frac32\delta^2\lambda^{\frac23}
-\frac{\delta^4}{\nu^2}+\frac32B\delta^2\lambda^{\frac43},~\delta\in\R. %
\label{eq:h}
\end{equation}
Recall that $\theta >0$ because of our assumption that $\mu>r$. Set 
\begin{eqnarray}
M&=&\theta+1+\frac2{-pA}\max\left\{6\frac{\sigma^2}{\nu}\left(2\nu\theta\abs{(1-\theta)(1-2\theta)} +1\right)+1,\right.
\label{eq:M}\\
&&\qquad \left.\frac12\sigma^2(1-p)\nu^2\max\limits_{t\in\zeroClosedTclosed}\{\xi(t)\} +1,1\right\}.\nonumber
\end{eqnarray}

Additionally, for $(t,\delta)\in\zeroClosedTclosed\times\R$,  we define functions
\begin{eqnarray}
f_1^{\pm}(t,\delta)  &\define&
\nu\lambda-p\nu\gamma_2(t) e^{-pA(T-t)}\lambda^{\frac53} \pm
p\nu M e^{-pA(T-t)}\lambda^2-ph(\delta)\lambda \nonumber\\
&&\qquad
+\Bigl(1+(\theta+\delta)\lambda\Bigr)h'(\delta),\label{eq:f1}\\
f_2^{\pm}(t,\delta)  &\define&
\nu\lambda-p\nu\gamma_2(t) e^{-pA(T-t)}\lambda^{\frac53}
\pm p\nu M e^{-pA(T-t)}\lambda^2-p h(\delta)\lambda\nonumber\\
&&\qquad
+\Bigl(-1+(\theta+\delta)\lambda\Bigr)h'(\delta).
\label{eq:f2}
\end{eqnarray}

\begin{lemma}
\label{lemma:roots}
For $t\in \zeroClosedTclosed$, there are continuous functions
\begin{equation}
\delta_1^{\pm}(t)\define-\frac12\nu\lambda^{\frac13}(1-\xi(t)\lambda^{\frac13})
+o\bigl(\lambda^{\frac23}\bigr),\quad
\delta_2^{\pm}(t)\define\frac12\nu\lambda^{\frac13}(1-\xi(t)\lambda^{\frac13})
+o\bigl(\lambda^{\frac23}\bigr)
\label{eq:delta-def}
\end{equation}
satisfying $f_i^{\pm}(t,\delta_i^{\pm}(t))=0$, $i=1,2$. For $t\in \zeroClosedTclosed $, $\delta_1^{\pm}$ and $\delta_2^{\pm}$ are also twice differentiable. In \eqref{eq:delta-def} the terms $o\bigl(\lambda^{\frac23}\bigr)$ are uniform in $t\in\zeroClosedTclosed$ consistent with Remark \ref{remark:order}.
\end{lemma}

{\sc Proof:}
The proof is given in the appendix.

\begin{Def}\label{Def:regions}
Choose $\lambda>0$ small enough that $\zeta^{\pm}_1(t) \triangleq
\theta +\delta_1^{\pm}(t)$ and $\zeta_2^{\pm}(t) \triangleq \theta +
\delta_2^{\pm}(t)$ all lie in $(0,1/\lambda)$. (We have $\theta >
0$ since $\mu > r$.) Define the {\em``no-trade"} region  %
$\NTpm\define\left\{(t,z) \Big|, t\in\zeroClosedTopen,~z\in(\zeta^{\pm}_1(t),\zeta^{\pm}_2(t))\right\},$ the buy region $\Bpm\define \left\{(t,z) \Big|,  t\in\zeroClosedTopen,~ -\frac{1}{\lambda}<z <\zeta^{\pm}_1(t)\right\}$, and analogously the sell region $\Spm\define \left\{(t,z) \Big|,  t\in\zeroClosedTopen,~ \frac{1}{\lambda}>z >\zeta^{\pm}_2(t)\right\}.$

\end{Def}

\begin{remark}\label{remark:bounds}
For $\lambda$ small enough, it follows from Definition \ref{Def:regions} and Lemma \ref{lemma:roots} that for $(t,z)\in\overline\NTpm$
\begin{equation}
\abs{  z-\theta }\le \nu\lambda^{\frac13},
\label{eq:delta-bnds}
\end{equation}
and we conclude that for $(t,z)\in\overline\NTpm$ 
\begin{equation}
\abs{h(z-\theta)} \le h_{\max}\lambda^{\frac43},~~\abs{h'(z-\theta)} \le \tilde h_{\max}\lambda,~~\abs{h''(z-\theta)} \le\tilde {\tilde {h}}_{\max}\lambda^{\frac23}.\label{eq:h-bnds}
\end{equation}

\end{remark}

\subsection{ \emph{{Step 2: Construction of the functions $w^{\pm}$}}} 

Define
\begin{equation}
w^{\pm}(t,z) \define
\begin{cases} \Bigl(\frac{1}{p} e^{pA(T-t)} - \gamma_2(t)\lambda^\frac{2}3 \pm M\lambda- \frac{e^{pA(T-t)}}{\nu} h(\zeta_1^{\pm}(t)-\theta)\Bigr)   \Bigl(\frac{1+\lambda z}{1+         \lambda \zeta_1^{\pm}(t)}   \Bigr)^{p}, \\
\qquad\qquad\qquad\qquad\qquad\qquad\qquad t\in\zeroClosedTopen,~ -1/\lambda \le z < \zeta_1^{\pm}(t)\\
\frac{1}{p} e^{pA(T-t)} - \gamma_2(t)\lambda^\frac{2}3 \pm M\lambda- \frac{e^{pA(T-t)}}{\nu} h(z-\theta),\\
\qquad\qquad\qquad\qquad\qquad\qquad\qquad t\in\zeroClosedTopen,~ \zeta_1^{\pm}(t) \le z\le \zeta_2^{\pm}(t)\\
 \Bigl(\frac{1}{p} e^{pA(T-t)} - \gamma_2(t)\lambda^\frac{2}3 \pm M\lambda- \frac{e^{pA(T-t)}}{\nu} h(\zeta_2^{\pm}(t)-\theta) \Bigr) \, \Bigl(\frac{1-\lambda             z}{1-\lambda \zeta_2^{\pm}(t)}\Bigr)^{p},\\
\qquad\qquad\qquad\qquad\qquad\qquad\qquad t\in\zeroClosedTopen,~ \zeta_2^{\pm}(t) < z \le 1/ \lambda.
\end{cases}
\label{eq:w}
\end{equation}
As a reminder, we have agreed to treat $w^{\pm}(t,\pm\frac{1}{\lambda})\triangleq-\infty, ~t\in\zeroClosedTclosed$, when $p<0$. Also note that if $M$ and $B$ were zero and $\gamma_3(t)$ were ignored, then in the $\NTpm$ region the formula for $w^{\pm}(t,z)$ agrees with the power series expansion \eqref{eq:heuristic-u}.
The term $\pm M\lambda$ in the definition of $w^{\pm}$ will be used to create the inequalities $\pm\mathcal H w^{\pm}\ge0$. Outside of the $\NTpm$ region, we extend this definition so that  $w^{\pm}$ would satisfy 
\begin{eqnarray}
w^{\pm}(t,z) &=& w^{\pm}(t,\zeta_1^{\pm}(t))\left(\frac{1+\lambda z}{1+\lambda\zeta_1^{\pm}(t)}
\right)^{p}, \quad t\in\zeroClosedTclosed,~ -\frac{1}{\lambda}\le z\leq \zeta_1^{\pm}(t),\label{4.1}\\
w^{\pm}(t,z) &=& w^{\pm}(t,\zeta_2^{\pm}(t))\left(\frac{1-\lambda z}{1-\lambda \zeta_2^{\pm}(t)}
\right)^{p},\quad t\in\zeroClosedTclosed,~ \zeta_2^{\pm}(t)\leq z \le \frac{1}{\lambda}. \label{4.2}
\end{eqnarray}
 We then have the derivative formula for $t\in\zeroClosedTclosed$,
\begin{equation}
w_z^{\pm}(t,z) = 
\begin{cases}
\frac{\lambda p}{1+\lambda z}w^{\pm}(t,z),& \! \! \!-\frac{1}\lambda<z< \zeta_1^{\pm}(t),\\
- \frac{e^{pA(T-t)}}{\nu} h'(z-\theta),& \zeta_1^{\pm}(t) < z < \zeta_2^{\pm}(t), \\
-\frac{\lambda p}{1-\lambda z} w^{\pm}(t,z),& \zeta_2^{\pm}(t) <z <\frac{1}\lambda.
\end{cases}
\label{eq:w_z}
\end{equation}
\begin{remark}\label{remark:C1}
The extensions \eqref{4.1}  and \eqref{4.2} ensure that the operators $\mathcal B,\mathcal S$ from \eqref{eq:Bu} and \eqref{eq:Su} satisfy $\mathcal B(w^{\pm})=0$ and $\mathcal S(w^{\pm})=0$ for $(t,z)\in \Bpm$ and $(t,z)\in \Spm$, respectively. Moreover, the equations $f_1^{\pm}(t,\delta_1^{\pm}(t)) = 0$ and $f_2^{\pm}(t,\delta_2^{\pm}(t)) =0$ guarantee that $w_z^{\pm}$ is defined and continuous at $(t,\zeta_1^{\pm}(t))$ and $(t,\zeta_2^{\pm}(t))$ for $t\in\zeroClosedTopen$. %
\end{remark}

We also have for $t\in\zeroClosedTclosed$
\begin{equation}
w_{zz}^{\pm}(t,z) = \begin{cases}
-\frac{\lambda^2 p(1-p)}{(1 + \lambda z)^2} w^{\pm}(t,z),&\! \! \!-\frac{1}\lambda <z <\zeta_1^{\pm}(t),\\
- \frac{e^{pA(T-t)}}{\nu}h''(z-\theta),& \zeta_1^{\pm}(t)< z<\zeta_2^{\pm}(t), \\
-\frac{\lambda^2 p(1-p)}{(1-\lambda z)^2}w^{\pm}(z),\quad&\zeta_2^{\pm}(t)<z <\frac{1}\lambda.
\end{cases}
\label{eq:w_zz}
\end{equation}
The function $w^{\pm}(t,z)$ is twice differentiable with respect to $z$ except on the curves $(t,\zeta_1^{\pm}(t))$ and $(t,\zeta_2^{\pm}(t)),~t\in\zeroClosedTopen$, where one-sided second derivatives with respect to $z$ exist and equal the respective one-sided limits of the
second derivatives.

For $(t,z)\in\Bpm$ we use \eqref{4.1} to calculate the derivatives with respect to time to be
\begin{eqnarray}
&&\!\!\!\!\!\!\!\!\!\!\!\!w_t^{\pm}(t,z)=\left(\frac{1+\lambda z}{1+\lambda\zeta_1^{\pm}(t)}
\right)^{p} \label{eq:w_t1}\\
&&\times \left( w^{\pm}_t(t,\zeta_1^{\pm}(t))+\left[w^{\pm}_z(t,\zeta_1^{\pm}(t))- w^{\pm}(t,\zeta_1^{\pm}(t))\frac{\lambda p}{1+\lambda\zeta_1^{\pm}(t)} \right]\frac{\du\zeta_1^{\pm}(t)  }{\dt} \right)\nonumber\\
&&=\left(\frac{1+\lambda z}{1+\lambda\zeta_1^{\pm}(t)}\right)^{p} w^{\pm}_t(t,\zeta_1^{\pm}(t)),\nonumber
\end{eqnarray}
where in the last equality we have used the fact that $(\mathcal B w^{\pm}) (t,\zeta_1^{\pm}(t))=0$. Indeed $\mathcal B$ defined in \eqref{eq:Bu} satisfies $\mathcal B(w^{\pm}) =0$ on $\Bpm$. The desired result follows because of continuous differentiability of $w^{\pm}(t,z)$ with respect to $z$.
Similarly for $(t,z)\in \Spm$, the derivatives with respect to time is
\begin{equation}
w_t^{\pm}(t,z)=\left(\frac{1-\lambda z}{1-\lambda\zeta_2^{\pm}(t)}
\right)^{p}w^{\pm}_t(t,\zeta_2^{\pm}(t)) .
\label{eq:w_t3}
\end{equation}
Finally for $(t,z)\in\NTpm$ we have that
\begin{equation}
w_t^{\pm}(t,z) =-pA (w^{\pm}(t,z) \mp M\lambda)  + \gamma_2e^{pA(T-t)} \lambda^{\frac23}.
\label{eq:w_t2}
\end{equation}

\begin{remark}\label{remark:C11}
As before, we see that $w^{\pm}(t,z)$ is  differentiable with respect to $t$ except on the curves $(t,\zeta_1^{\pm}(t))$ and $(t,\zeta_2^{\pm}(t)),~t\in\zeroClosedTopen$, where one-sided derivatives exists and equal the respective one-sided limit of the derivatives. Together with Remark \ref{remark:C1} we conclude that $w^{\pm}\in C^{1,1}(\zeroClosedTopen\times\Su).$
\end{remark}

\subsection{ \emph{{Step 3: Verification that $\mathcal H w^{-} \le 0$}}} \label{step3}

Recall the operators $\mathcal H, \mathcal D, \mathcal B,\mathcal S$ from \eqref{eq:HJB2}, \eqref{eq:Du}, \eqref{eq:Bu} and \eqref{eq:Su} respectively. It suffices to verify
\begin{equation}\label{Subsolution}
-w_t^{-}(t,z) +\mathcal D w^{-}(t,z) \leq 0,\quad t\in\zeroClosedTopen,~
\zeta_1^-(t)<z<\zeta_2^-(t).
\end{equation}
We, thereby, simultaneously also develop an analogous inequality for $w^+$ needed in the subsequent section. %
Therefore,
\begin{eqnarray*}
\lefteqn{-w_t^{\pm}(t,z) +\mathcal D w^{\pm}(t,z)}\\
&=& pA\left( w^{\pm}(t,z) \mp M\lambda \right) - \gamma_2 e^{pA(T-t)} \lambda^{\frac23}-p\Bigl(A -\frac{1}2 \sigma^2(1-p)(z-\theta)^2\Bigr)w^{\pm}(t,z)\\
&& -(1-p)\sigma^2z(1-z)(z-\theta) \frac{e^{pA(T-t)}}{\nu}h'(z-\theta)+\frac{1}2 \sigma^2
z^2(1-z)^2 \frac{e^{pA(T-t)}}{\nu}h''(z-\theta)\\
&=& \mp pA M\lambda - \gamma_2 e^{pA(T-t)}\lambda^{\frac23}+\frac{p}2 \sigma^2(1-p)(z-\theta)^2w^{\pm}(t,z) \\
&&-(1-p)\sigma^2z(1-z)(z-\theta) \frac{e^{pA(T-t)}}{\nu}h'(z-\theta)\\
&&+\frac12 \sigma^2z^2(1-z)^2 \frac{e^{pA(T-t)}}{\nu}\left[3 \lambda^{2/3}-\frac{12(z-\theta)^2}{\nu^2} +3B \lambda^{4/3}\right].
\end{eqnarray*}
Writing $z=\theta+(z-\theta)$ and
$1-z=1-\theta-(z-\theta)$, and using \eqref{eq:delta-bnds} we compute
\begin{eqnarray}
z^2(1-z)^2&=&\theta^2(1-\theta)^2
+O\bigl(\lambda^{\frac13}\bigr),\label{eq:z-lambda13}\\
\abs{z^2(1-z)^2-\theta^2(1-\theta)^2} &=& 2\abs{(z-\theta)\theta(1-\theta)(1-2\theta)} +O(\lambda^{\frac23})\nonumber\\
&\le&\left[2\nu\theta\abs{(1-\theta)(1-2\theta)} +1\right]\lambda^\frac13,\label{eq:z-lambda23}
\end{eqnarray}
where the last inequality holds for $\lambda$ small enough. From Remark \ref{remark:bounds} we obtain
\begin{eqnarray}
\lefteqn{-w_t^{\pm}(t,z) +\mathcal D w^{\pm}(t,z)}\label{eq:inside-NT}\\
&=& \left[-\gamma_2+\frac32 \frac{\theta^2(1-\theta)^2\sigma^2}{\nu}\right] e^{pA(T-t)} \lambda^{\frac23}\nonumber\\
&&+\frac12\left[(1-p)-\frac{12\theta^2(1-\theta)^2}{\nu^3}\right](z-\theta)^2\sigma^2e^{pA(T-t)}\nonumber\\
&&\mp pA M\lambda \left( 1- \frac{1}{2A} \sigma^2(1-p)(z-\theta)^2\right) +O(\lambda).\nonumber
\end{eqnarray}
The definitions of $\gamma_2$ and $\nu$ imply that the first two terms on the right-hand side are zero. For $\lambda$ small enough using \eqref{eq:delta-bnds}, we have that $ 1 - \frac{1}{2A} \sigma^2(1-p)(z-\theta)^2\ge \frac12.$ 
We conclude that by the definition \eqref{eq:M}  of $M$ equation \eqref{eq:inside-NT} can be made positive (negative), since using \eqref{eq:z-lambda23} it can be shown that the $O(\lambda)$ term above can be bounded by $\left[6\frac{\sigma^2}{\nu}\left(2\nu\theta\abs{(1-\theta)(1-2\theta)} +1\right)+1\right] \lambda.$
Here we have used our assumption that $pA<0$, see also Remark \ref{remark:beta}. We conclude that in $\NTm$ we have $0 \ge -w_t^{-}(t,z) +\mathcal D w^{-}(t,z) \ge \mathcal H (w^{-})(t,z)$. 

This completes the verification that $\mathcal H (w^{-}) \le 0$ in $\zeroClosedTopen\times \Su$, since $w^{-}\in C^{1,1}(\zeroClosedTopen\times\Su)$, and in the buy $\Bm$ region we have by Remark \ref{remark:C1} that $\mathcal H (w^{-}) \le \mathcal B (w^{-})=0$, and a similar inequality holds in the sell region $\Sm$.

$\hfill\Box$

\subsection{ \emph{{Step 4: Verification that $\mathcal H (w^+) \ge 0$}}}
\subsubsection{ \emph{{Step 4a: Verification that $\mathcal H (w^+) \ge 0$ in $\Bp$}}} \label{step4a}

By construction we have that $(\mathcal B w^+)(t,z)=0$ for $(t,z)\in\Bp.$ Since $pw^+(t,z) \ge 0,$ it follows that $w_z^+(t,z) \ge 0,$ and we conclude that $(\mathcal S w^+) (t,z) \ge 0$ there.
It remains to verify that for sufficiently small $\lambda$
\begin{equation}-w_t^+(t,z) +(\mathcal D w^+)(t,z) \geq 0,\quad (t,z)\in \Bp.
\label{eq:in1} 
\end{equation}
Using \eqref{4.1}, \eqref{eq:w_z}, \eqref{eq:w_zz} and \eqref{eq:w_t1}, we conclude that
\begin{eqnarray}
&&\!\!\!\!\!\!\!\!\!-w_t^+(t,z) +(\mathcal D w^+)(t,z)=\left(\frac{1+\lambda z} {1+\lambda\zeta_1^{\pm}(t)}\right)^{p}\label{eq:in2} \\
&&\times\left[  -w^{+}_t(t,\zeta_1^{\pm}(t)) -p\Bigl(A -\frac{1}2 \sigma^2(1-p)(z-\theta)^2\Bigr)w^{+}(t,\zeta_1^{+}(t))\right.\nonumber\\
&& \quad  +(1-p)\sigma^2z(1-z)(z-\theta) \frac{\lambda p}{1+\lambda z}w^{+}(t,\zeta_1^{+}(t))\nonumber\\
&&\quad \left.+\frac{1}2 \sigma^2
z^2(1-z)^2 \frac{\lambda^2 p(1-p)}{(1 + \lambda z)^2}w^{+}(t,\zeta_1^{+}(t))\right].\nonumber
\end{eqnarray}

For a fixed $t$, it is easy to verify that for $\lambda>0$ sufficiently small, the function $k(z)\triangleq(z-\theta)+\lambda z(1-z)/(1+\lambda z)$ attains its maximum over
$(-1/\lambda, \zeta_1^+(t)]$ at $\zeta_1^+(t)$ and
$k(\zeta_1^+(t)) < 0$. 
Therefore for $(t,z)$ such that $-\frac{1}\lambda < z \leq \zeta_1^+(t)$
\begin{equation*}
k^2(z) \geq k^2(\zeta_1^+(t))= (\delta_1^+(t))^2 + O(\lambda) =\frac{\nu^2}4\lambda^{\frac23}-\frac{\nu^2}2\xi(t)\lambda +o\bigl(\lambda\bigr),
\end{equation*} 
and we conclude that for $\lambda$ small enough
\begin{equation}
k^2(z) \geq  \frac{\nu^2}4\lambda^{\frac23}-\nu^2\xi(t)\lambda.
\label{eq:in3}
\end{equation} 
It follows from \eqref{eq:in2} and \eqref{eq:in3} that for sufficiently small $\lambda>0$
\begin{eqnarray*}
&&\!\!\!\!\!\!\!\!\!\left(\frac{1+\lambda z} {1+\lambda\zeta_1^{+}(t)}\right)^{-p}\left[-w_t^+(t,z) +(\mathcal D w^+)(t,z)\right]=\\
&&  pA (w^{+}(t,\zeta_1^{+}(t)) - M\lambda)  - \gamma_2 e^{pA(T-t)}\lambda^{\frac23} -pAw^{+}(t,\zeta_1^{+}(t))\\
&&\quad +\frac12 \sigma^2 p(1-p) k^2(z)w^{+}(t,\zeta_1^{+}(t))\\
&&\ge -pAM\lambda -\gamma_2e^{pA(T-t)}\lambda^\frac23+\frac12\sigma^2p(1-p)\left[\frac14\nu^2\lambda^{\frac23}-\nu^2\xi(t)\lambda \right]\\
&&\qquad\times\left[\frac{1}{p} e^{pA(T-t)} - \gamma_2(t)\lambda^\frac{2}3 + M\lambda-\frac{e^{pA(T-t)}}{\nu} h(\zeta_1^{+}(t)-\theta)\right]\\
&&=+e^{pA(T-t)}\left[-\gamma_2+\frac18\sigma^2(1-p)\nu^2\right] \lambda^{\frac23}-      \frac12\sigma^2(1-p)\nu^2\xi(t)e^{pA(T-t)}\lambda\\
&&\quad -pAM\lambda\left(  1- \frac{\nu^2}{2A}\sigma^2(1-p)\left[\frac14\lambda^{\frac23}-\xi(t)\lambda\right]  \right) + O(\lambda^{\frac43}),%
\end{eqnarray*}
where the term $O(\lambda^{\frac43})$ can be shown to be $-\frac{1}2 \sigma^2p(1-p) \nu\left[\frac14\lambda^{\frac23}-\xi(t)\lambda \right] \left[\nu\gamma_2(t)\lambda^\frac{2}3 \right.$ $\left.+  e^{pA(T-t) } h(\zeta_1^{+}(t)-\theta)  \right]$ and hence can be bounded by $\lambda$, for $\lambda$ small enough.
By definitions of $\nu$ and $\gamma_2$ the $\lambda^{\frac23}$ term is zero. Moreover, for $\lambda$ small enough, we have that $1- \frac{\nu^2}{2A}\sigma^2(1-p)\left[\frac14\lambda^{\frac23}-\xi(t)\lambda\right]  \ge \frac12$. We conclude that \eqref{eq:in1} holds for $M$ satisfying \eqref{eq:M}.  
\begin{remark}\label{remark:w^-}
A similar calculation shows that for any $t\in \zeroClosedTopen$ 
$$\lim%
\limits_{\begin{array}{c}
(s,z)\to(t,\zeta_1^-(t))\\
(s,z)\in \Bp, s\in\zeroClosedTopen
\end{array}}
-w_t^-(s,z) +(\mathcal D w^-)(s,z) \le 0.$$
\end{remark}

$\hfill\Box$

\subsubsection{ \emph{{Step 4b: Verification that $\mathcal H (w^+) \ge 0$ in $\Sp$}}} \label{step4b}

This is analogous to  \emph{Step 4a}.\\

\subsubsection{ \emph{{Step 4c: Verification that $\mathcal H (w^+) \ge 0$ in $\NTp$}}} \label{step4c}
In \emph{ Step 3} we have shown that for $(t,z)\in \NTp$
$$
-w_t^+(t,z)+\mathcal (D w^+)(t,z)\geq 0.
$$
We must also show that for $(t,z) \in \overline\NT^+$,
\begin{equation}\label{eq:g}
g(t,z)\define
\lambda pw^+(t,z)-(1+\lambda z)w_z^+(t,z) \geq
0.%
\end{equation}
Fix $t\in \zeroClosedTopen$. For $z\in[\zeta_1^+(t),\zeta_2^+(t)]$, we have $z-\theta= O\bigl(\lambda^{1/3}\bigr)$. Using this fact, we compute
\begin{eqnarray*}
g_z(t,z)&=&-w_{zz}^+(t,z)+O(\lambda^\frac53)=\frac{e^{pA(T-t)}}{\nu}h''(z-\theta)+O(\lambda^\frac53)\\
&=& \frac{12}{\nu}e^{pA(T-t)}\lambda^{\frac23}\left[\frac14- \left(\frac{z-\theta}{\nu\lambda^{\frac13}}\right)^2+O(\lambda)\right].
\end{eqnarray*}
We know that $g(t,\zeta_1^+(t))=0$ and thus, to prove (\ref{eq:g}), it suffices to show for our fixed $t$ that $g_z(t,\cdot)$ is positive on $[\zeta_1^+(t),\zeta_2^+(t)]$. Because $-(z-\theta)^2$ is a concave function of $z$, it suffices to check the endpoints. We have for $i=1,2$ that
$$
\left(\frac{\zeta_i^+(t)-\theta}{\nu\lambda^{\frac13}}\right)^2
=\left(\pm\frac12(1-\xi(t)\lambda^{\frac13}) +o\bigl(\lambda^{\frac13}\bigr)\right)^2
=\frac14-\frac12\xi(t)\lambda^{\frac13}+o\bigl(\lambda^{\frac13}\bigr).
$$
Therefore,
$$
g_z(t,\zeta_i^+(t))= \frac{12}{\nu}e^{pA(T-t)}\lambda^{\frac23}
\left[\frac12\xi(t)\lambda^{\frac13}+o\bigl(\lambda^{\frac13}\bigr)\right]>0
$$
for sufficiently small $\lambda>0$ independently of $t$. The proof that
$$h(t,z)\define\lambda pw^+(t,z)+(1-\lambda z)w_z^+(t,z),~(t,z) \in \overline\NT^+$$
is positive for $z\in[\zeta_1^+(t),\zeta_2^+(t)]$ is analogous. 
This completes the verification that $\mathcal H (w^+) \ge 0$ on $\zeroClosedTopen\times \Su.$ 

\par

So far we have constructed two continuous differentiable functions $w^{\pm}(t,z) \in %
C^{1,1}(\zeroClosedTopen\times \Su)$ and showed that they satisfy $\pm (\mathcal H)(w^{\pm}) \ge 0$ on $\zeroClosedTopen\times \Su.$ By definition $w^{\pm}$ satisfy the boundary condition $w^{\pm}(t,z)=0$ if $0<p<1$ and $w^{\pm}(t,z)=-\infty$ if $p<0$ for $(t,z)\in \zeroClosedTclosed\times \partial \Su$, and we have that $w^{\pm}(t,\theta) = \frac{1}{p} e^{pA(T-t)} -\gamma_2(t)\lambda^\frac{2}3+O(\lambda),~t\in\zeroClosedTclosed$. To conclude the proof of Theorem \ref{Thm1} we are left only to verify the final time conditions.

$\hfill\Box$

\subsection{ \emph{{Step 5: Final time conditions}}}

For $(T,z)$ such that $\zeta_1^{+}(T)\le z \le\zeta_2^{+}(T)$, from \eqref{eq:w} %
for $\lambda$ small enough we have
$$ w^{+}(T,z)=\frac{1}{p}+M\lambda-\frac{h(z-\theta)}{\nu}>\frac{(1-\lambda z)^p}{p}=u(T,z),
$$
since $\frac{(1-\lambda z)^p}{p}=\frac1p-\lambda z+O(\lambda^2)$ and because of Remark \ref{remark:bounds}.

When $\frac1\lambda > z >\zeta_2^{+}(T)$ from \eqref{4.2} we see that
\begin{eqnarray*}
&& w^{+}(T,z)=w^{+}(T,\zeta_2^{+}(T))\left(\frac{1-\lambda z} {1-\lambda\zeta_2^{+}(T)}\right)^{p} > u(T,\zeta_2^{+}(T))\left(\frac{1-\lambda z}{1-\lambda \zeta_2^{+}(T)}
\right)^{p}=u(T,z).
\end{eqnarray*}
Next, consider the case when $0 \le z <\zeta_1^{+}(T)$. For $\lambda$ small enough, from \eqref{4.1} we have

\begin{eqnarray*} 
&&w^{+}(T,z)=w^{+}(T,\zeta_1^{+}(T))\left(\frac{1+\lambda z} {1+\lambda\zeta_1^{+}(T)}\right)^{p} \\
&&=\left(\frac{1+\lambda z} {1+\lambda\zeta_1^{+}(T)}\right)^{p}\left(\frac{1}{p}+M\lambda\right)+ O(\lambda^{\frac43})\\
&&= \left(1-\lambda p(\zeta_1^{+}(T)-z) + O(\lambda^2)\right)\left(\frac{1}{p}+ M\lambda\right)+ O(\lambda^{\frac43}) %
>\frac{(1-\lambda z)^p}{p}=u(T,z),
\end{eqnarray*}
where the inequality follows because for $\lambda$ small enough $(1-\lambda p(\zeta_1^{+}(T)-z)+O(\lambda^2)) \ge \frac12$, $M$ satisfies \eqref{eq:M} and all the $O(\lambda^{\frac43})$ are uniform in $z\in[0,\zeta_1^{+}(T))$ consistent with Remark \ref{remark:order}.

We now see that when $-\frac1\lambda < z <0$, from \eqref{4.1} it follows that
\begin{eqnarray*} 
&&w^{+}(T,z)=w^{+}(T,\zeta_1^{+}(T))\left(\frac{1+\lambda z} {1+\lambda\zeta_1^{+}(T)}\right)^{p}=w^{+}(T,0)\left(1+\lambda z\right)^{p} \\
&&> u(T,0)\left(1+\lambda z\right)^{p}=u(T,z).
\end{eqnarray*}
Analogously, we see that $w^{-}(T,z)<u(T,z)$.%

This completes the proof of Theorem \ref{Thm1}.

$\hfill\Box$
\section{``Nearly-Optimal" Strategy}\label{sec:nearly-optimal_strat}
In this section we will show that the {\em``no-trade"}, buy and sell regions $\NTu, \Bu$ and $\Slu$ from Definition \ref{Def:regions} and the strategy associated with these regions is a ``nearly-optimal" strategy; see Theorem \ref{thm:main}. 
For $(t,x,y)\in\zeroClosedTopen\times\Sv,$ define the {\em``no-trade"} region in the original variables as $(t,x,y)\in \NTv \Leftrightarrow \left(t,\frac{y}{x+y}\right)\in\NTu.$ Similarly we define the buy and sell regions  $\Bv, \Slv$ in the original variables. 
\begin{lemma}\label{lemma:existence}
Let $t\in\zeroClosedTopen$ and $(x,y)\in\Sv$, and let $(\tilde L,\tilde M)$ be the strategy  associated with the {\em``no-trade"}, buy and sell regions $\NTu, \Bu$ and $\Slu$. Then there exists a strong solution $(\X_s,\Y_s)\big|_{s\in\tClosedTclosed}$ to \eqref{eq:position1} and \eqref{eq:position2}, such that $(\X_{t-},\Y_{t-})=(x,y).$  

\end{lemma}
{\sc Proof:}
We define the strategy $(\tilde L,\tilde M)$ to be the trading strategy associated with with the {\em``no-trade"}, buy and sell regions $\NTu, \Bu$ and $\Slu$. This strategy requires trading anytime the position is inside the buy or sell regions until the position reaches the boundary of the {\em``no-trade"} region. Then the strategy calls for buying (respectively selling) stock whenever the position is on the boundary of $\Bu$ (respectively $\Slu$), so that agent's position does not leave $\overline\NTu$. On the boundaries of the {\em``no-trade"} region these trades increase $\tilde L$ or $\tilde M$ and push the diffusion $(s,\X_s,\Y_s)$ in direction pointing to the inside of $\NTu$. We refer to these directions as (oblique) directions of reflection.

Note that this strategy is not optimal. It requires the agent to buy stocks, if she has a positive number of stocks but is still in the buy region, even at time $t=T$, regardless of the fact that to compute the final utility she would have to convert her stock position into cash. However, this causes loss of $O(\lambda)$, and we are able to prove that this is a ``nearly-optimal" strategy.
   
Define $\mathbb G_n\define \NTv \cap \left\{(s,x,y) | \frac1n<\abs{x}+\abs{y}<n\right\},$ and assume for convenience that $t=0$. 
The directions of reflection on $\partial \mathbb G_n\cap\partial \NTv$ are the same as on $\partial \NTv$, as long as we stop the process at\\ %
$\tau_n\define T\wedge\inf\left\{0<s\le T: \abs{\X_s}+\abs{\Y_s} \le \frac1n \mbox{ or } \abs{\X_s}+\abs{\Y_s} \ge n\right\},$ so we treat the other two boundaries $\left\{(s,x,y): \abs{x}+\abs{y}=n\right\}$ and $\left\{(s,x,y):    \abs{x}+\abs{y}=\frac1n\right\}$ as absorbing. The reader can verify that \emph{Case 1} conditions of \emph{Theorem 4.8} of  Dupuis \& Ishii \cite{DupuisIshii} are satisfied on $[0,\tau_n),$ which gives us the existence of the processes $\X_s,\Y_s$  and the local time process $\tilde L_s,\tilde M_s$ on $[0,\tau_n).$ 
Indeed there are two condition for \emph{Case 1}. The first condition requires that there is a unique direction of reflection $r(s,x,y)$ on the boundary and that it changes smoothly as a function of a point of the boundary $(s,x,y)$. In our case, the directions of reflection are $(0, -(1+\lambda),1)$ and $(0,1-\lambda, -1)$ on the buy and sell boundaries $\partial\Bv\cap\partial\mathbb G_n , \partial\Slv\cap\partial\mathbb G_n$ respectively. The second condition requires that $\exists b\in(0,1)$ such that $\cup_{0\le t\le b} \mbox{Ball}_{tb}( (s,x,y) - tr(s,x,y)) \subset \mathbb G_n^{c}$ for $(s,x,y)\in \partial\mathbb G_n\cap\partial\NTv,$ where $\mbox{Ball}_{r}(z)$ is a ball of radius $r$ centered at $z$. In our case, it is easy to see that both conditions are satisfied.

Letting $n\rightarrow \infty$ we get $\X_s,\Y_s,\tilde L_s$ and $\tilde M_s$ on $[0,\tau)$, for $\tau\define\lim\limits_{n\rightarrow\infty}\tau_n.$ %
Define $\tilde L_\tau\define\lim\limits_{s\to\tau} \tilde L_s,$ and analogously $\tilde M_\tau.$ Note that $\tilde L_\tau,\tilde M_\tau$ exist.
The only thing left to verify is that $\X_s,\Y_s$ are semi-martingales, that is, $\tilde L_{\tau}, \tilde M_{\tau}$ are finite a.s. Assume the opposite, that is that on some set $\mathbf A$ of positive probability at least one of them, say $\tilde L_{\tau}=\infty.$ Consider the processes
\begin{eqnarray*}
 \du \hat X_t &=& r \hat X_t\dt - \,\du \tilde L_t + \,\du
 \tilde M_t, \qquad \qquad \quad X_{0-}=x,\\
 \du \hat Y_t &=& \mu \hat Y_t \dt + \sigma \hat Y_t \,\du W_t + \du\tilde  L_t - \du
 \tilde M_t,~~~Y_{0-}=y.%
\end{eqnarray*}
that correspond to the wealth invested in the money market and in stock respectively using the same strategy $(\tilde L, \tilde M)$, but in a market without transaction costs. It follows that
$\hat Y_{\tau_n}=\Y_{\tau_n},$ and
$$\hat X_{\tau_n}=\e{r\tau_n} x_0-\int_0^{\tau_n} \e{r(\tau_n-s)} d\tilde L_s +\int_0^{\tau_n} \e{r(\tau_n-s)} d\tilde M_s.$$
Analogously
$$\X_{\tau_n}=\e{r\tau_n}x_0-(1+\lambda)\int_0^{\tau_n} \e{r(\tau_n-s)}d\tilde L_s +(1-\lambda)\int_0^{\tau_n} \e{r(\tau_n-s)}d\tilde M_s.$$
It follows that because $\X_s+\Y_s \ge 0$ for all $s\in\zeroClosedTopen$ and $\tilde L,\tilde M$ are increasing
\begin{equation}
\hat X_{\tau_n}+\hat Y_{\tau_n} =  \X_{\tau_n}+\Y_{\tau_n}+ \lambda\int_0^{\tau_n} \e{r(\tau_n-s)}\left(d\tilde L_s +d\tilde M_s\right) \ge \lambda  \left( \tilde L_{\tau_n}+ \tilde M_{\tau_n}\right).
\label{eq:Xhat+Yhat}
\end{equation}
Fix $N \in \mathbb{N}$ a big integer. Then, there exists $n$ such that $\tilde L_{\tau_n}>N$ on a set $\tilde{\mathbf A}$, such that $\tilde{\mathbf A}\subset {\mathbf A}$ and $\P\left(\tilde{\mathbf A}\right)>\P({\mathbf A}) -\frac1N$. For that $n$ consider the strategy $\hat L,\hat M$ in the zero-transaction cost model that is  identical to $\tilde L,\tilde M$ on the set $[0,\tau_n)$ and sells all the stock at time $\tau_n$, that is, $\hat L_s\define\hat L_{\tau_n}$ and $\hat M_s\define\hat M_{\tau_n}+\hat Y_{\tau_n}$ for $s \in [\tau_n,T].$ We have from \eqref{eq:Xhat+Yhat}
$$\E\Big[\U\left(\lambda\big(\hat L_{\tau_n}+\hat M_{\tau_n}\big)\right) \ind_{\tilde{\mathbf A}}   \Big]\le\E\Big[\U\left(\hat X_{\tau_n}+\hat Y_{\tau_n}\right)  \ind_{\tilde{\mathbf A}}   \Big]\le \E\Big[ \U\left(\hat X_T+\hat Y_T\right) \ind_{\tilde{\mathbf A}}   \Big]. $$
In the case of $0<p<1$, this leads to a contradiction since the left hand side is at least $\U(\lambda N) \P \left(\tilde{\mathbf A}\right),$ whereas the right hand side is bounded by $\frac{1}{p}e^{pA(T-t)}(x+y)^{p}$, the value function for the case of zero transaction costs. If $p<0$ then we can apply the above argument with $\tilde p =\frac12$, since the admissibility of the strategy $(\tilde L_s,\tilde M_s)\big|_{s\in\tClosedTclosed}$ is independent of $p$.

We finish the construction by defining $\X_\tau=\lim\limits_{s\to\tau}\X_s$ and analogously $\Y_\tau$. On the set $\left\{\tau<T\right\}$ note that $\X_\tau=\Y_\tau=0$ a.s. because of the following Remark \ref{remark:boundedness}. For $s\in(\tau,T]$ we define $\X_s\define\Y_s\define\X_{\tau}=\Y_{\tau}=0,$ and $\tilde L_s\define\tilde L_{\tau}, \tilde M_s\define\tilde M_{\tau}$. %

$\hfill\Box$

\begin{remark}\label{remark:boundedness}
Similar to the argument above, it is easily shown that for any $(t,x,y)\in\zeroClosedTclosed\times\Sv$ and any admissible strategy $( L_s, M_s)\big|_{s\in\tClosedTclosed} \in \mathcal A(t,x,y)$ and for any stopping time $\tau$ satisfying $\tau\in\tClosedTclosed,$ we have that $\P \left(\left\{X_\tau+Y_\tau<\infty \right\}\right)=1.$

\end{remark}

\begin{theorem}\label{thm:comp}
Assume $p<1,~p\ne0,~pA<0~,\theta>0,~\theta\ne1$ and $\lambda >0.$  Let $w^{\pm}$ be the functions constructed in Theorem \ref{Thm1}. Then $w^{+}(t,z)\geq u(t,z)$ and $w^{-}(t,z)\leq u(t,z)$ for $(t,z)\in\zeroClosedTclosed\times\overline\Su$.
\end{theorem}

{\sc Proof:}  
If $t=T$ or $z \in \partial \Su$ then the claim follows from Theorem \ref{Thm1} and Lemma \ref{lemma:zero_boundary}. For $(t,z)\in\zeroClosedTOpen\times\Su$, let $\psi^{\pm}(t,x,y)\define(x+y)^{p}w^{\pm}\big(t,y/(x+y)\big),~(x,y)\in\Sv.$ Consider the upper bound case first. In light of (\ref{eq:valueFct}), it suffices to prove
that $\psi^{+}(t,x_0,y_0)\geq v(t,x_0,y_0)$ for
fixed but arbitrary $(t,x_0,y_0)\in\zeroClosedTOpen\times\Sv$. Let $(L_s,M_s)\big|_{s\in\tClosedTclosed}\in\mathcal A(t,x_0,y_0)$ be an admissible policy for this initial position $(t,x_0,y_0)$. The  function $\psi^{+}$ is of class $C^{1,2,2}$ in $\zeroClosedTopen\times\Sv$ except possibly on the curves $(s,x,y)$, where $(s,y/(x+y))=(s,\zeta_i^{+}(s)),~s\in\tClosedTopen,~i=1,2$, where $\psi^{+}$ is $C^{1,1,1}.$ 

Define $\tau_n\define T\wedge \inf\{t \le s \le T; |X_s+Y_s-\lambda|Y_s||\leq 1/n, \abs{X_s}+\abs{Y_s} \ge n\}$, and let $\tau\define\lim\limits_{n\rightarrow\infty}\tau_n$. We can mollify $\psi^{+}$ to obtain $\psi^{+}_\varepsilon\define\psi^{+}*\varphi_\varepsilon$ a $C^{1,2,2}$ function, where $\varphi_\varepsilon$ is a standard mollifier. Apply It\^o's rule to $\psi^{+}_\varepsilon$ to get 
\begin{eqnarray}
&&\!\!\!\!\!\!\!\!\!\!\!\!e^{-\beta \tau_n} \psi^{+}_\varepsilon(\tau_n,X_{\tau_n},Y_{\tau_n})-e^{-\beta t}\psi^{+}_\varepsilon(t,X_t,Y_t)  = -\int_t^{\tau_n}e^{-\beta s} 
\Bigl[ -(\psi^{+}_\varepsilon)_t(s,X_s,Y_s)\,\du s  \nonumber\\
&& \quad + \mathcal L \psi^{+}_\varepsilon(s,X_s,Y_s)\,\du s+ \big((1+\lambda)(\psi^{+}_\varepsilon)_x(s,X_s,Y_s)  - (\psi^{+}_\varepsilon)_y(s,X_s,Y_s)
\big)\,\du L_s  \nonumber\\
&& \quad +\big(-(1-\lambda)(\psi^{+}_\varepsilon)_x(s,X_s,Y_s) +(\psi^{+}_\varepsilon)_y(s,X_s,Y_s)\big)\,\du M_s\Bigr] \nonumber\\
&& + \sigma\int_t^{\tau_n}e^{-\beta s}Y_s(\psi^{+}_\varepsilon)_y(s,X_s,Y_s) \,\du W_s. %
\label{4.5}
\end{eqnarray}

Since $\psi^{+}$ is $C^{1,1,1},$ then the limits as $\varepsilon \downarrow 0$ of $\psi^{+}_\varepsilon$ and of the first derivatives of $\psi^{+}_\varepsilon$ are respectively $\psi^{+}$ and the appropriate first derivatives of $\psi^{+}$. Then the limit as $\varepsilon\searrow0$ of $\int_t^{\tau_n} e^{-\beta s}Y_s^2(\psi^{+}_{\varepsilon})_{yy}(s,X_s,Y_s)ds$ exists, since from \eqref{4.5} it can be expressed using $\psi^{+}$ and its first derivatives. By the dominated convergence theorem we have 
\begin{eqnarray}
&&\!\!\!\!\!\!\!\!\!\!\!\!\int_t^{\tau_n} e^{-\beta s}Y_s^2\underline{ \psi^{+}_{yy}}(s,X_s,Y_s)ds \le 
 \lim\limits_{\varepsilon\searrow 0 }\int_t^{\tau_n}e^{-\beta s}Y_s^2 (\psi^{+}_{\varepsilon})_{yy}(s,X_s,Y_s)ds ~~\label{4.5.1}\\
&&\le \int_t^{\tau_n}e^{-\beta s}Y_s^2 \overline{\psi^{+}_{yy}}(s,X_s,Y_s)ds,\nonumber
\end{eqnarray}
where $\overline{\psi^{+}_{yy}}(t_1,x_1,y_1)= \limsup\limits_{ (s,x,y)\rightarrow (t_1,x_1,y_1) }\psi^{+}_{yy}(s,x,y)$, and
\newline
$\underline{\psi^{+}_{yy}}(t_1,x_1,y_1)=\liminf\limits_{ (s,x,y)\rightarrow (t_1,x_1,y_1)}\psi^{+}_{yy}(s,x,y).$

From Theorem \ref{Thm1}, the dominated convergence theorem and \eqref{4.5.1} it follows that
\newline
$\lim\limits_{\varepsilon\searrow 0 }-\int_t^{\tau_n}e^{-\beta s} 
\Bigl[ -(\psi^{+}_\varepsilon)_t(s,X_s,Y_s)+ \mathcal L \psi^{+}_\varepsilon(s,X_s,Y_s) \Bigl]\,\du s \le 0.$ Passing to the limit as $\varepsilon\searrow 0$ in \eqref{4.5} we conclude that
\begin{equation}
e^{-\beta\tau_n}\psi^{+}(\tau_n,X_{\tau_n},Y_{\tau_n})
\le e^{-\beta t}\psi^{+}(t,x_0,y_0)+\sigma\int_t^{\tau_n} e^{-\beta s}Y_s\psi^{+}_y(s,X_s,Y_s)\,\du W_s.\label{4.7}
\end{equation}
Consider first the case $0<p<1$. In this case 
\begin{equation}
e^{-\beta T}\U(X_T+Y_T-\lambda|Y_T|) \le e^{-\beta\tau}\psi^{+}(\tau,X_\tau,Y_\tau),
\label{4.7.1}
\end{equation}
because it holds on the set $\{\tau=T\}$, and on the set $\{\tau<T\}$ $X_{\tau}+Y_{\tau}-\lambda|Y_{\tau}|=0$ a.s. It follows from Lemma \ref{lemma:zero_boundary} that $X_T+Y_T-\lambda|Y_T|=0$ a.s. there too and both sides of \eqref{4.7.1} are zero. Take expectation of both sides of \eqref{4.7}. Then by Fatou's Lemma and using \eqref{4.7.1} we conclude that
\begin{equation}
\E\left[e^{-\beta T}\U(X_T+Y_T-\lambda|Y_T|)\right] 
\le e^{-\beta t}\psi^{+}(t,x_0,y_0).\label{4.8}
\end{equation}
Taking the supremum over all admissible policies, we conclude that $\psi^{+}(t,x_0,y_0) \ge v(t,x_0,y_0).$

We now consider the case $p<0$. For this case, we need further to assume that $(L_s,M_s)\big|_{s\in\tClosedTclosed}$ is not just any admissible strategy, but the optimal strategy, the existence of which is shown, for example, in Dai \& Yi \cite{DaiYi}. We then have that $v(t,x_0,y_0)=\E\left[e^{-\beta (T-t)}\U(X_T+Y_T-\lambda|Y_T|)\right]$. 

If $\E\left[\U(X_T+Y_T-\lambda|Y_T|)\right] =-\infty$ then \eqref{4.8} is trivially true, and $\psi^{+}(t,x_0,y_0) \ge v(t,x_0,y_0)$. Otherwise, $\P\left(\{X_T+Y_T-\lambda|Y_T| >0\}\right)=1$. Under this assumption, we show that 
\begin{equation}
\big\{\tau=T\big\}=\bigcup^\infty_{n=1}\big\{\tau_n=T\big\} \mbox{ a.s.}
\label{4.8.1}
\end{equation}
It is clear that $\{\tau=T\}$ contains the union. For the reverse containment, assume that $\omega \not\in \bigcup^\infty_{n=1}\big\{\tau_n=T\big\}$. From Remark \ref{remark:l.s.c} $X_{\cdot}(\omega)+Y_{\cdot}(\omega)-\lambda\abs{Y_{\cdot}(\omega)}$ is lower semi-continuous function on $[t,T]$, it then follows from Remark \ref{remark:boundedness} that $X_{\tau}(\omega)+Y_{\tau}(\omega)-\lambda\abs{Y_{\tau}(\omega)}=0$ a.s. and by Lemma \ref{lemma:zero_boundary} we have that $X_T(\omega)+Y_T(\omega)-\lambda|Y_T(\omega)|=0$ and we are on a set of measure zero, and \eqref{4.8.1} follows. We conclude that $\lim\limits_{n\to \infty } \P\left(\big\{\tau_n=T\big\}\right)=1$.
Take expectation of both sides of \eqref{4.7} to get
\begin{equation} \E\left[e^{-\beta \tau_n}\psi^{+}(\tau_n,X_{\tau_n},Y_{\tau_n})\left(\ind_{\{\tau_n=T\}}+\ind_{\{\tau_n<T\}}\right)\right]\le e^{-\beta t}\psi^{+}(t,x_0,y_0),%
\label{4.8.3}
\end{equation}
where we have used that $\tau_n\le T$ by definition. We also have
\begin{equation}
\E\left[e^{-\beta T}\U(X_T+Y_T-\lambda|Y_T|)\ind_{\{\tau_n=T\}}\right] \le \E\left[e^{-\beta \tau_n}\psi^{+}(\tau_n,X_{\tau_n},Y_{\tau_n})\ind_{\{\tau_n=T\}}\right].
\label{4.8.4}
\end{equation}
The left hand side of \eqref{4.8.4} converge by the monotone convergence theorem to\\  $e^{-\beta T}\E\left[\U(X_T+Y_T-\lambda|Y_T|)\right]$. To conclude that \eqref{4.8} holds, it is enough to show that 
$$\lim\limits_{n\to \infty }\E\left[\psi^{+}(\tau_n,X_{\tau_n},Y_{\tau_n})\ind_{\{\tau_n<T\}}\right]=0.$$
Dai \& Yi \cite{DaiYi} show that the ratio $\frac{Y_s}{X_s+Y_s}$ is in some compact set $\mathbb K \subset\R$, when $(X_s,Y_s)\ne(0,0)$. Indeed, Dai \& Yi \cite{DaiYi} show that $Y_s \ge 0$, i.e. that it is never optimal to short stock, due to our assumption that $\mu>r$. In addition in case of $(\mu-r) - (1-p)\sigma^2 \le 0$, in Remark 4.6, they also show that $\frac{X_s}{Y_s} \ge 0$, when $(X_s,Y_s)\ne(0,0)$. It follows that $0\le \frac{Y_s}{X_s+Y_s}\le 1$ then.
Additionally, in case  $(\mu-r) - (1-p)\sigma^2 >  0$, in Remark 6.2, they also show that  $\frac{X_s}{Y_s} \ge \left( \frac{(1-p)\sigma^2}{2(\mu-r)-(1-p)\sigma^2}-1 \right) (1-\lambda)$. Then for $\lambda \in (0,1)$ 
$$0\le \frac{Y_s}{X_s+Y_s}\le \frac1{\frac{(1-p)\sigma^2}{2(\mu-r)-(1-p)\sigma^2}(1-\lambda) + \lambda} \le \frac{2(\mu-r)}{(1-p)\sigma^2}-1.$$

Assume $\lambda$ is small enough, so that $\mathbb K\subset\Su$.
From the boundedness of $u$ and $w^{+}$ on $\zeroClosedTclosed\times\mathbb K$ and the definitions of $v$ and $\psi^{+}$, we conclude that there exists a constant $c>0$, such that $\psi^{+}(s,x,y) \ge c v(s,x,y),$ whenever $y/(x+y)$ is in $\mathbb K$ or $(x,y)=(0,0)$. 
It follows that 
\begin{eqnarray*}
&&0\ge\E\left[\psi^{+}(\tau_n,X_{\tau_n},Y_{\tau_n})\ind_{\{\tau_n<T\}}\right]\ge
c\E\left[v(\tau_n,X_{\tau_n},Y_{\tau_n})\ind_{\{\tau_n<T\}}\right]\ge\\
&&c\E\left[\E\big[e^{-\beta (T-\tau_n)}\U(X_T+Y_T-\lambda|Y_T|)\big|\mathcal F_{\tau_n}\big]\ind_{\{\tau_n<T\}}  \right] \rightarrow 0,
\end{eqnarray*}
because $\lim\limits_{n\to \infty } \P\left(\big\{\tau_n<T\big\}\right)=0$, and $\U(X_T+Y_T-\lambda|Y_T|)$ is integrable. We conclude that 

\begin{equation}
v(t,x_0,y_0)=\E\left[e^{-\beta(T-t)}\U(X_T+Y_T-\lambda|Y_T|)\right] 
\le \psi^{+}(t,x_0,y_0).\label{4.8.2}
\end{equation}

Now we treat the lower bound case. We show for fixed but arbitrary $(t,x_0,y_0)\in\zeroClosedTOpen\times\Sv$ that $\psi^{-}(t,x_0,y_0)\leq v(t,x_0,y_0)$. This time consider the admissible policy $(\tilde L,\tilde M)$ that was constructed in Lemma \ref{lemma:existence}. 
The diffusion $(\X_s,\Y_s)\big|_{s\in\tClosedTopen}$ spends Lebesgue-measure zero time on the boundaries of $\NTv$.

Consider first the case when $p<0$. Applying It\^o's formula to mollifications of $\psi^{-}$ we get an equality similar to \eqref{4.5}. Note that the integrals with respect to $d\tilde L_s$ and $d\tilde M_s$ are zero by construction of the subsolution $w^{-}$. From Theorem \ref{Thm1} passing to the limit as $\varepsilon\searrow 0$ we get the reverse inequality in (\ref{4.7}). We also have the reverse inequality in \eqref{4.7.1}. Take expectation, and again use Fatou's Lemma to conclude that

\begin{eqnarray}
&&\!\!\!\!\!\!\!\!\!\!\!\!\!\!\! e^{-\beta t}v(t,x_0,y_0) \ge \E\left[e^{-\beta T}\U(\X_T+\Y_T-\lambda|\Y_T|)\right] %
\label{4.9}\\
&&\!\!\!\!\!\!\!\!\!\!\!\!\!\!\! \ge  \E\left[e^{-\beta \tau}\psi^{-}(\tau,\X_\tau,\Y_\tau)\right]\ge e^{-\beta t}\psi^{-}(t,x_0,y_0).\nonumber
\end{eqnarray}
In case $0<p<1$, let $\tau_n\define T\wedge \inf\{t \le s \le T; \psi^{-}(s,\X_s,\Y_s)\leq 1/n, \abs{X_s}+\abs{Y_s} \ge n\},~\nu_n\define T\wedge \inf\{t \le s \le T; \abs{X_s}+\abs{Y_s} \ge n\}$ and $\tau\define\lim\limits_{n\rightarrow\infty}\tau_n.$ %
We repeat the above argument and get a reverse inequality in \eqref{4.7}, and taking expectation we conclude that
$$\E\left[e^{-\beta\tau_n}\psi^{-}(\tau_n,\X_{\tau_n},\Y_{\tau_n}) \left(\ind_{\{\tau_n=T\}}+\ind_{\{\tau_n<T\}}\right)\right] \ge e^{-\beta t}\psi^{-}(t,x_0,y_0).$$
We have that
\begin{eqnarray*}
&&\E\left[e^{-\beta\tau_n}\psi^{-}(\tau_n,\X_{\tau_n},\Y_{\tau_n}) \ind_{\{\tau_n=T\}}\right] \le \E\left[e^{-\beta T}\U(\X_T+\Y_T-\lambda|\Y_T|) \ind_{\{\tau_n=T\}}\right]\\
&&\le\E\left[e^{-\beta T}\U(\X_T+\Y_T-\lambda|\Y_T|) \right].
\end{eqnarray*}
Since $\frac{\Y_s}{\X_s+\Y_s}$ is in some compact set $\mathbb K \subset\R$, when $(\X_s,\Y_s)\ne(0,0)$, and assume that $\lambda$ is small enough so that $\mathbb K\subset\Su$. Then from the boundedness of $u$ and $w^{-}$ on $\zeroClosedTclosed\times\mathbb K$ and the definitions of $v$ and $\psi^{-}$ we conclude that there exists a constant $c>0$, such that $\psi^{-}(s,x,y) \le c v(s,x,y),$ whenever $y/(x+y)$ is in $\mathbb K$ or $(x,y)=(0,0)$. From Remark \ref{remark:boundedness} $\P\left( \tau_n= \nu_n<T\right) \rightarrow0$. Similar to the argument above we conclude that $\E\left[e^{-\beta\tau_n}\psi^{-}(\tau_n,\X_{\tau_n},\Y_{\tau_n}) \ind_{\{\tau_n= \nu_n<T \}}\right]$ converges to zero. Moreover, since $\psi^{-}(\tau_n,\X_{\tau_n},\Y_{\tau_n}) \ind_{\{\tau_n<T\wedge \nu_n \}} \le \frac1n$ then
$$\E\left[e^{-\beta\tau_n}\psi^{-}(\tau_n,\X_{\tau_n},\Y_{\tau_n}) \ind_{\{\tau_n<T\wedge \nu_n \}}\right]\rightarrow 0.$$ 
It follows that
$$\E\left[e^{-\beta\tau_n}\psi^{-}(\tau_n,\X_{\tau_n},\Y_{\tau_n}) \ind_{\{\tau_n<T\}}\right]\rightarrow 0.$$ 
We conclude that
\begin{equation}
\E\left[e^{-\beta T}\U(\X_T+\Y_T-\lambda|\Y_T|)\right] \ge e^{-\beta t } \psi^{-}(t,x_0,y_0).\
\label{4.10}
\end{equation}

$\hfill\Box$

\begin{remark}\label{remark:nearly-optim-strat}
Define the expected discounted utility of the final wealth associated with the ``nearly-optimal" strategy,
\begin{eqnarray}
&&\!\!\!\!\!\!\!\!\!\!\!\!\!\!\!\!\!\!\!\!\!\!\!\tilde v(t,x,y)\define \E \left[e^{-\beta(T-t)} \U\left(\X_T+\Y_T-\lambda\abs{\Y_T}\right)\right],~ (t,x,y)\in\zeroClosedTclosed\times\overline\Sv, \label{eq:v-bar}\\
&&\!\!\!\!\!\!\!\!\!\!\!\!\!\!\!\!\!\!\!\!\!\!\!\tilde u(t, z) \define \bar v(t,1-z,z),\qquad (t,z) \in \zeroClosedTclosed\times\overline\Su.
\label{eq:u-bar}
\end{eqnarray}
It follows from \eqref{4.9} and \eqref{4.10} that for $(t,x,y) \in \zeroClosedTclosed\times \overline \Sv$ we have that $\psi^{-}(t,x,y) \le \tilde v (t,x,y)$ or equivalently $w^{-}(t,z)\le \tilde u(t,z)$ for $(t,z) \in \zeroClosedTclosed\times \overline \Su.$
\end{remark}

We can now finally prove Theorem \ref{thm:main}.

\par

{\sc Proof of Theorem \ref{thm:main}:}
From Theorem \ref{thm:comp} we see that $w^{-}(t,z) \le u(t,z)\le w^{+}(t,z).$ Moreover, $w^+(t,z) - w^-(t,z) = O(\lambda)$, and $w^\pm(t,z) = w^\pm(t,\theta) + O(\lambda)$ for fixed $(t,z)\in\zeroClosedTclosed\times\Kone$. It follows that
$u(t,z) = w^\pm(t,z) + O(\lambda) = w^\pm(t,\theta) + O(\lambda) = u(t,\theta)+O(\lambda)$.

Moreover, from Remark \ref{remark:nearly-optim-strat} we have that $w^{-}(t,z)\le \tilde u(t,z)$ for $(t,z) \in \zeroClosedTclosed\times \overline \Su.$ We conclude that the strategy $(\tilde L, \tilde M)$ from Lemma \ref{lemma:existence} is ``nearly-optimal", that is it matches the value function of the optimal strategy up to order $O(\lambda^{\frac23})$.

$\hfill\Box$

\section{Appendix}

{\sc Proof of Lemma \ref{lemma:roots}:}
We shall only consider $\delta$ of order $O\bigl(\lambda^{1/3}\bigr)$. For such $\delta$, $h(\delta)=O(\lambda^\frac43)$, $h'(\delta)=O(\lambda)$, so it follows from (\ref{eq:f1}) that for $t\in\zeroClosedTclosed$ we have
\begin{eqnarray}
\lefteqn{f_1^\pm(t,\delta)}\label{eq:f1-tmp1}\\
&=& \nu\lambda -p\gamma_2(t)\nu e^{-pA(T-t)} \lambda^{\frac53}+ h'(\delta)+ O\bigl(\lambda^2\bigr)\nonumber\\
&=&\nu\lambda-p\gamma_2(t)\nu e^{-pA(T-t)}\lambda^\frac53 + 3\delta\lambda^\frac23- \frac{4\delta^3}{\nu^2} +3B\delta\lambda^{\frac43}+ O\bigl(\lambda^2\bigr).\nonumber
\end{eqnarray}
Consider $\delta_0 \triangleq -\frac{1}2 \nu \lambda^{1/3}
\left(1 - \xi_0(t) \lambda^{1/3} \right) ,$ where $\xi_0(t) \define \sqrt{\xi^2(t) + \etaT}$, with $\etaT$ satisfying $\abs{\etaT} < \min\limits_{s\in\zeroClosedTclosed}\xi^2(s)$.  From \eqref{eq:gamma-zeta} it follows that
\begin{equation} \begin{split}
f_1^\pm(t,\delta_0) = \:& \nu\lambda -p(T-t)\gamma_2\nu \lambda^\frac53 - \frac32 \nu \lambda +\frac32 \nu \xi_0(t) \lambda^\frac43-\frac32B\nu\lambda^{\frac53} \\
& + \frac{1}2 \nu\,\lambda \Bigl(1-3 \xi_0(t)\lambda^\frac{1}3
+ 3 \xi_0^2(t) \lambda^\frac23\Bigr) + O\bigl(\lambda^2\bigr)\\
= \:& \nu \Bigl(\frac{3}2 \xi_0^2(t)-p(T-t)\gamma_2-\frac32B\Bigr) \lambda^\frac53 + O\bigl(\lambda^2\bigr).
\end{split}
\label{eq:f1-tmp2}
\end{equation}

Then $f_1^\pm(t,\delta_0) = \frac32 \nu \etaT\lambda^\frac53 + O\bigl(\lambda^2\bigr)$. Thus, when $\etaT > 0$ we have $f_1^\pm(t,\delta_0)>0$, and when $\etaT < 0$ we have $f_1^\pm(t,\delta_0)<0$ for sufficiently small $\lambda>0$. Therefore, for $t\in \zeroClosedTclosed$, and any $\etaT \in \bigl(0,\min\limits_{s\in\zeroClosedTclosed}\xi^2(s)\bigr)$ for sufficiently small $\lambda>0$ there exists
$$
\delta_1^\pm(t) \in \left(-\frac12 \nu \lambda^\frac13\Bigl(1-\lambda^\frac13 \sqrt{\xi^2(t) - \etaT}\Bigr),-\frac12 \nu \lambda^\frac13\Bigl(1-\lambda^\frac13 \sqrt{\xi^2(t) +\etaT}\Bigr) \right)
$$
satisfying $f_1^\pm(t,\delta_1^\pm(t)) = 0$. In other words, $\delta_1^\pm(t) = -\frac{1}2 \nu \lambda^\frac13\,\Bigl(1-\xi(t)\lambda^\frac13\Bigr) + o\bigl(\lambda^\frac23\bigr)$. 
\\
The proof of the existence of $\delta_2^\pm(t)$ is analogous.\newline
For $t\in \zeroClosedTclosed,$  we note that for $\delta=-\frac12\nu\lambda^{\frac13}\Bigl(1-\xi(t)\lambda^\frac13\Bigr)+o(\lambda^{\frac23})$ we have
$ \frac{\partial f_1^\pm(t,\delta)}{\partial t} = p\nu\gamma_2\lambda^{\frac53} +O(\lambda^2),$ 
and  
$ \frac{\partial f_1^\pm(t,\delta)}{\partial \delta} = 3\lambda^{\frac23} -\frac{12 \delta^2}{\nu^2}+ O(\lambda^{\frac43})=6\xi(t)\lambda+o(\lambda).$
By the implicit function theorem $\delta_1^\pm(t)$ is a continuously differentiable function and its derivative is
\begin{equation}
\frac{\du \delta_1^\pm(t)}{\dt} = -\frac{\frac{\partial f_1^\pm(t,\delta)}{\partial t} }{\frac{\partial f_1^\pm(t,\delta)}{\partial \delta}}=O(\lambda^{\frac23})
\label{eq:delta_t}
\end{equation}
The second derivative can be computed similarly.

$\hfill\Box$

\end{document}